\documentclass[11pt,a4]{article}
\usepackage{amsmath}
\usepackage{amscd}
\usepackage{array}
\usepackage{latexsym,amssymb}
\usepackage{caption}
\usepackage{epsfig}
\newcommand{\ft}[2]{{\textstyle\frac{#1}{#2}}}
\DeclareFontFamily{U}{rsf}{}
\DeclareFontShape{U}{rsf}{m}{n}{
  <5> <6> rsfs5 <7> <8> <9> rsfs7 <10-> rsfs10}{}
\DeclareMathAlphabet\Scr{U}{rsf}{m}{n}

\usepackage{color}

 \csname
@addtoreset\endcsname{equation}{section}

\newcommand{\nn}{\nonumber}

\newcommand{\be}{\begin{equation}}
\newcommand{\ee}{\end{equation}}
\newcommand{\bea}{\begin{eqnarray}}
\newcommand{\eea}{\end{eqnarray}}

\newcommand{\ba}{\begin{eqnarray}}
\newcommand{\ea}{\end{eqnarray}}

\def\notin{\hbox{{$\in$}\kern-.51em\hbox{/}}}

\def\inbar{\vrule height1.5ex width.4pt depth0pt}
\def\IB{\relax{\rm I\kern-.18em B}}
\def\IC{\relax\,\hbox{$\inbar\kern-.3em{\rm C}$}}
\def\ID{\relax{\rm I\kern-.18em D}}
\def\IE{\relax{\rm I\kern-.18em E}}
\def\IF{\relax{\rm I\kern-.18em F}}
\def\IG{\relax\,\hbox{$\inbar\kern-.3em{\rm G}$}}
\def\IH{\relax{\rm I\kern-.18em H}}
\def\II{\relax{\rm I\kern-.17em I}}
\def\IK{\relax{\rm I\kern-.18em K}}
\def\IL{\relax{\rm I\kern-.18em L}}
\def\IN{\relax{\rm I\kern-.18em N}}
\def\IP{\relax{\rm I\kern-.18em P}}
\def\IQ{\relax\,\hbox{$\inbar\kern-.3em{\rm Q}$}}
\def\IR{\relax{\rm I\kern-.18em R}}
\def\IU{\relax\,\hbox{$\inbar\kern-.3em{\rm U}$}}
\def\ZZ{\relax\ifmmode\mathchoice{\hbox{\cmss
Z\kern-.4em Z}}{\hbox{\cmss Z\kern-.4em Z}}{\lower.9pt\hbox{\cmsss Z\kern-.4em Z}} {\lower1.2pt\hbox{\cmsss
Z\kern-.4em Z}}\else{\cmss Z\kern-.4em Z}\fi}
\def\IGam{\relax{{\rm I}\kern-.18em \Gamma}}

\def\bfnull{\relax{\rm O \kern-.635em 0}}

\def\square{{\,\lower0.9pt\vbox{\hrule
\hbox{\vrule height 0.2 cm \hskip 0.2 cm \vrule height 0.2 cm}\hrule}\,}}

\def\twomat#1#2#3#4{\left(\begin{array}{cc} \end{array} \right)}

\begin{document}

\begin{titlepage}
\rightline{}
%\rightline{\tt }
\rightline\today
\begin{center}
\vskip .6cm
{\LARGE \bf {The ${\cal N}>2$ supersymmetric AdS vacua\\[1.2ex]
in maximal supergravity}}\\
\vskip 1.0cm
{\large\bf {Antonio Gallerati${\,}^1$, Henning Samtleben${\,}^2$, Mario Trigiante${\,}^1$}}
\vskip .6cm
{\it ${}^1$ Dipartimento di Fisica, Politecnico di Torino\\
C.~so Duca degli Abruzzi, 24, I-10129 Torino, Italy} \\
antonio.gallerati@polito.it,\,mario.trigiante@polito.it

\vskip 0.2cm
{\it {${}^2$ Universit\'e de Lyon, Laboratoire de Physique, UMR 5672, CNRS}}\\
{\it {\'Ecole Normale Sup\'erieure de Lyon}}\\
{\it {46, all\'ee d'Italie, F-69364 Lyon cedex 07, France}}\\
henning.samtleben@ens-lyon.fr

\vskip .5cm
{\bf Abstract}

\end{center}

We perform a systematic search for anti-de Sitter vacua of maximal supergravity
with \mbox{$\mathcal{N}>2$} residual supersymmetries.
We find that maximal supergravity admits two 1-parameter classes of
$\mathcal{N}=3$ and $\mathcal{N}=4$ vacua, respectively.
They are embedded, for the different values of an angular parameter, in the $\omega$-rotated
${\rm SO}(8)$ ($\mathcal{N}=3$) and ${\rm SO}(1,7)$ ($\mathcal{N}=4$ and 3)  gauged models.
All vacua disappear in the $\omega\rightarrow 0$ limit. We determine the mass spectra and the
AdS-supermultiplet structure. These appear to be the first and only $\mathcal{N}>2$ supersymmetric AdS
vacua in maximal supergravity, aside from the $\mathcal{N}=8$ vacua of the ${\rm SO}(8)$-gauged models.
We also prove on general grounds that no such vacua can exist for $4<\mathcal{N}<8$.
\vskip 0.2cm

\noindent
\begin{narrower}

\end{narrower}

\end{titlepage}

\newpage

\section{Introduction}
Since the seminal paper \cite{deWit:1982ig} in which the first gauged maximal supergravity was constructed with gauge group ${\rm SO}(8)$, much work has been done to study the vacua of this model and to construct new gauged maximal supergravities.
Certain vacua of the original ${\rm SO}(8)$-gauged model\footnote{See \cite{Warner:1983vz} for early results in the search for vacua of the original theory and \cite{Fischbacher:2008zu} for a recent study.}, like the anti-de Sitter (AdS) vacuum with $\mathcal{N}=8$ residual supersymmetry, were put in correspondence with compactifications of $D=11$ supergravity on a seven-dimensional sphere or on warped/stretched versions of a seven-sphere, possibly with torsion. Non-compact and even non-semisimple gaugings, defined by groups of the form ${\rm CSO}(p,q,r)$, $p+q+r=8$, were first constructed in \cite{Hull:1984qz} and their de Sitter vacua put in correspondence with reductions on non-compact spaces with negative curvature \cite{Hull:1988jw}. Flat-gaugings in $D=4$ describing Scherk-Schwarz reductions of maximal $D=5$ supergravity  and yielding no-scale models, were first constructed in \cite{Andrianopoli:2002mf}.\par
A new formulation of gauged extended supergravities in terms of the so called \emph{embedding tensor} \cite{deWit:2002vt,deWit:2005ub,deWit:2007mt}, has opened the way for a more systematic analysis of the possible gaugings and their vacuum structure. All possible choices of gauge groups in the maximal supergravity are encoded in a single object $\Theta_M{}^\alpha$ (the embedding tensor), which defines the embedding of the gauge algebra inside the algebra $\mathfrak{e}_{7(7)}$ of the on-shell global symmetry group ${\rm E}_{7(7)}$ of the ungauged theory. This object  is formally ${\rm E}_{7(7)}$--covariant and is constrained, by linear and quadratic conditions originating from the  requirement of supersymmetry and gauge invariance, to belong to certain orbits of the ${\bf 912}$ representation. An interesting feature of this formulation is that the field equations and Bianchi identities of the gauged model are formally ${\rm E}_{7(7)}$-covariant if the fields are transformed together with the embedding tensor. In other words there is  a mapping (or duality) between gauged theories defined by embedding tensors that are related by ${\rm E}_{7(7)}$ transformations. Such mapping should encode the effect of string/M-theory dualities on flux compactifications.
In particular the scalar potential $V(\Theta,\phi)$ is a quadratic function of $\Theta_M{}^\alpha$ and is invariant under the simultaneous action of ${\rm E}_{7(7)}$ on the $70$ scalar fields $\phi=(\phi^{ijkl})$  of the model and the embedding tensor:
\begin{equation}
\forall g\in {\rm E}_{7(7)}\,\,;\,\,\,\,V(\Theta,\phi)=V(g\star \Theta,g\star\phi)\,,
\end{equation}
where $g\star$ denotes the generic action of a group element $g$ on the scalars (non-linear action) and on $\Theta_M{}^\alpha$ (linear action).
The above property and the homogeneity of the scalar manifold has motivated what has been dubbed as the ``going-to-the-origin'' approach for the study of vacua of gauged supergravities \cite{Dibitetto:2011gm,DallAgata:2011aa}: Any vacuum of a given gauged model can be mapped into the origin of the scalar manifold\footnote{By origin we mean the point in the scalar coset manifold ${\rm E}_{7(7)}/{\rm SU}(8)$ at which all scalar fields $\phi^{ijkl}$ vanish and is thus manifestly ${\rm SU}(8)$-invariant.} by means of a suitable  ${\rm E}_{7(7)}$-transformation, provided the embedding tensor is transformed accordingly. This means that the vacua of gauged maximal supergravity can be systematically studied by restricting to the origin of the manifold so that the extremization condition on $V$ becomes another condition on  $\Theta_M{}^\alpha$ only. In this way, one can search for vacua with particular properties without committing to a particular gauge group, i.e.\ while simultaneously scanning through all possible gaugings.\par
In \cite{DallAgata:2012bb} a new family of ${\rm SO}(8)$-gauged maximal supergravities was constructed by exploiting the freedom in the original choice of the symplectic frame defining the electric and magnetic gauge fields. These models were obtained as a deformation of the original  de Wit and Nicolai model, parametrized by an angle $\omega$. They all exhibit an $\mathcal{N}=8$ vacuum at the origin. Their spectrum is identical while the $\omega$ parameter only affects the higher-order interactions. Similar generalizations  of non-compact gaugings were studied in \cite{DallAgata:2012sx,DallAgata:2014ita}. Adopting the ``going-to-the-origin'' approach, the authors of \cite{Borghese:2012qm,Borghese:2012zs,Borghese:2013dja} systematically searched for vacua with certain residual symmetries  and found several vacua of the new $\omega$-deformed models.
An interesting feature observed in all the above  works, is that the   $\omega$-deformed models in general exhibit a much richer vacuum structure that the original $\omega=0$ models from \cite{deWit:1982ig,Hull:1984qz}. In other words, many vacua of these theories disappear in the limit $\omega\rightarrow 0$.\par
In the present paper we start a systematic analysis of  vacua of maximal supergravity with a minimal amount of residual supersymmetry.
We focus on AdS vacua preserving $\mathcal{N}>2$ supersymmetries by implementing the supersymmetry conditions (Killing spinor equations) directly on the embedding tensor (with the scalar fields fixed at the origin).
Our (computer aided) analysis is systematic and we find, aside from the known $\mathcal{N}=8$ vacua, only two other classes of solutions with residual supersymmetry $\mathcal{N}=4$ and $\mathcal{N}=3$, respectively. \emph{These are, to our knowledge, the first AdS vacua of maximal supergravity with residual $\mathcal{N}>2$  supersymmetry, aside from the $\mathcal{N}=8$ ones.} We can exclude, by general argument, solutions with $8>\mathcal{N}>4$.
Each class of the newly found vacua is parametrized by an angle $\varphi$ and, depending on its values, the corresponding vacua are embedded in different ($\omega$-deformed) ${\rm CSO}(p,q,r)$ models. In particular the $\mathcal{N}=4$ vacua, depending on $\varphi$, belong to gaugings of the form ${\rm SO}(1,7)$ and $[{\rm SO}(1,1)\times {\rm SO}(6)]\ltimes T^{12}$, while $\mathcal{N}=3$ vacua to models with gauge group  ${\rm SO}(8)$, ${\rm SO}(1,7)$ and  ${\rm ISO}(7)$. \par
We compute the mass spectra on these vacua, which turn out to be $\varphi$-independent, and determine the corresponding  AdS-supermultiplet structure. Our analysis shows that, while there are several AdS ${\cal N}=8 \,\longrightarrow\, {\cal N}=3$  supersymmetry breaking patterns, only one, for each residual symmetry, seems to be dynamically realized in the full non-linear theory.\par
As a last comment, vacua with residual ${\rm SO}(4)$ symmetry were investigated in \cite{Borghese:2013dja}. This analysis however missed the vacua discussed here since it restricted the ${\rm SO}(8)$ singlets to a sector which is invariant under a $D_4$ discrete subgroup of ${\rm SU}(8)$.\footnote{As a consequence, the  gravitino mass matrix which is consistent with these symmetry requirements  is proportional to the identity matrix and thus is different from the one we obtain
for ${\cal N}=4$.}\par
The paper is organized as follows.
In Section \ref{sec:AdS} we formulate the problem of systematically studying the spontaneous $\mathcal{N}=8$ supersymmetry breaking on an AdS vacuum with residual extended supersymmetry: After a first introduction of the embedding tensor formalism, we consider the spontaneous supersymmetry breaking to $\mathcal{N}>2$ on AdS vacua and derive the corresponding system of quadratic equations on the non-vanishing components of the embedding tensor. We show that ${\cal N}>2$ residual supersymmetry requires the massive gravitinos to transform non-trivially under the associated ${\rm SO}({\cal N})$ R-symmetry group. In particular, we deduce the absence of solutions with $8>\mathcal{N}>4$ residual supersymmetry.\\
 In section \ref{N83}, we study the possible AdS ${\cal N}=8 \,\longrightarrow\, {\cal N}=3$  supersymmetry breaking patterns
 at the level of the corresponding supersymmetry multiplets.
In section \ref{sec:vacua}, we then describe the $\mathcal{N}=4$ and $\mathcal{N}=3$ classes of solutions to the quadratic equations. We identify, for the different values of the angular parameter $\varphi$, the corresponding gauge groups through the signature of the Cartan-Killing metric and by identifying the ${\rm E}_{7(7)}$-invariant quantities constructed out of the embedding tensor with the same quantities evaluated on $\omega$-rotated ${\rm SO}(8)$ \cite{DallAgata:2012bb},  ${\rm SO}(1,7)$ groups and on ${\rm ISO}(7)$.
 We show that these vacua disappear in the $\omega\rightarrow 0$ limit.
 Finally we give the AdS-supermultiplet structure and bosonic mass spectrum for the two classes of solutions.
Appendix~\ref{app:not} summarizes our conventions and normalizations for the mass matrices; appendix~\ref{app:details}
collects some of the computational details for the results of the main text.

\section{AdS vacua with extended supersymmetry}
\label{sec:AdS}

\subsection{Gauged ${\cal N}=8$ supergravity}

Let us briefly review some key formulas of gauged ${\cal N}=8$ supergravity,
for details we refer to~\cite{deWit:2002vt,deWit:2007mt,LeDiffon:2011wt}.
Gaugings of maximal ${\cal N}=8$ supergravity are described by the gauge group
generators $X_{MN}{}^K$, ($M, N = 1, \dots, 56$) which in turn are obtained by
contracting the ${\rm E}_{7(7)}$ generators $t_\alpha$ ($\alpha = 1, \dots, 133$) with a given embedding tensor
$\Theta_M{}^\alpha$
\bea
X_{MN}{}^K &=& \Theta_M{}^\alpha\,(t_\alpha)_N{}^K
\;.
\label{XMNK}
\eea
They satisfy the quadratic identity
\bea{}
[X_M, X_N] &=& -X_{MN}{}^K\,X_K
\qquad\Longleftrightarrow\qquad
\Omega^{MN} \Theta_M{}^\alpha \Theta_N{}^\beta ~=~ 0
\;,
\label{Xalgebra}
\eea
which poses a quadratic constraint on the embedding tensor $\Theta_M{}^\alpha$, and
exhibits the closure of the gauge algebra.
The dressing of the generators (\ref{XMNK}) with the scalar dependent complex vielbein
$\left\{{\cal V}_{M\,[ij]}, {\cal V}_M{}^{[ij]}\equiv({\cal V}_{M\,[ij]})^*\right\}$, $i, j =1, \dots, 8$, defines the $T$-tensor
\bea
(T_{ij})^{klmn} &\equiv&
\frac12\,({\cal V}^{-1})_{ij}{}^M ({\cal V}^{-1})^{kl}{}^N\,(X_{M})_{N}{}^K\,{\cal V}_K{}^{mn}
\;,\qquad
\mbox{etc.}
\;.
\label{TT}
\eea
The various components of this tensor will show up in the field equations
of the gauged theory and parametrize the couplings.
The fact that the embedding tensor $\Theta_M{}^\alpha$ is restricted to the ${\bf 912}$
representation of ${\rm E}_{7(7)}$ can be expressed by parametrizing the components of
the $T$-tensor according to
\bea
(T_{ij})_{kl}{}^{mn} &=&  \frac12\left(
\delta_{[k}{}^{[m} A{}^{n]}{}_{l]ij} +  \delta_{i[k}{}^{mn} A_{l]j}-\delta_{j[k}{}^{mn} A_{l]i}
\right)    \;,\nonumber\\[.5ex]
(T_{ij})^{rs}{}_{pq} &=& - \frac12\left(\delta_{[p}{}^{[r} A{}^{s]}{}_{q]ij} +  \delta_{i[p}{}^{rs} A_{q]j}-
 \delta_{j[p}{}^{rs} A_{q]i}\right) \;,\nonumber\\[.5ex]
(T_{ij})_{kl \, pq} &=& \frac{1}{24}\, \varepsilon_{klpqrstu} \, \delta_{[i}{}^r A_{j]}{}^{stu}
\;,\nonumber\\[.5ex]
(T_{ij})^{rs \; mn} &=& \delta_{[i}{}^{[r} A_{j]}{}^{smn]} \;,
\label{TAB}
\eea
in terms of the scalar tensors\footnote{Here and in the following the coupling constant $g$ is absorbed in the definition of the tensors  $A_{ij}$, $A_i{}^{jkl}$.} $A_{ij}$, $A_i{}^{jkl}$, satisfying
$A_{[ij]}=0$, $A_{i}{}^{jkl} = A_{i}{}^{[jkl]}$, and $A_{i}{}^{jki}=0$.
These tensors represent the ${\bf 36}$ and ${\bf 420}$
representations of ${\rm SU}(8)$, respectively, and
parametrize the Yukawa-type couplings in the Lagrangian as
\bea
{\cal L}_{\rm Yuk} &=&
 e\,\Big\{\frac12\sqrt{2}\,  A_{1\,ij}\,
\bar{\psi}^i_\mu
\gamma^{\mu\nu} \psi^j_\nu + \frac{1}{6}  A_{i}{}^{jkl} \,
\bar{\psi}^i_\mu \gamma^\mu \chi_{jkl}
\nonumber\\
&&{}\qquad\qquad
+
\frac1{144} \sqrt{2}\,
  \varepsilon^{ijkpqrlm}\,A{}^{n}{}_{pqr}
\,
\bar{\chi}_{ijk}\,\chi_{lmn} ~+~ {\rm h.c.}\Big \}
\;,
\label{Yuk}
\eea
for the eight gravitini $\psi_\mu{}^i$ and the 56 fermions $\chi_{ijk}$\,.
The quadratic constraints (\ref{Xalgebra}) on the embedding tensor induce the following
identities among the scalar dependent tensors $A_{ij}$, $A_i{}^{jkl}$
\begin{eqnarray}
 0&=&  A{}^k{}_{lij} \,A_{n}{}^{mij} - A_{l}{}^{kij} \, A{}^m{}_{nij}
  -4A{}^{(k}{}_{lni}A^{m)i}-4A_{(n}{}^{mki}A_{l)i} \nn\\
%%%%%%%%%%%%%
  &&{}-2\,\delta_{l}^{m}\,A_{ni}A{}^{ki}+2\,\delta_{n}^{k}\,A_{li}A{}^{mi}
 \;,\nonumber \\[1ex]
%%%%%%%%%%%%%
0&=&   A{}^i{}_{jk[m} \,A{}^k{}_{npq]}
  +A_{jk}\delta^{i}_{[m}A{}^{k}{}_{npq]}
  -A_{j[m}A^{i}{}_{npq]} \nn\\
 && {}+\frac1{24}\,\varepsilon_{mnpqrstu}\,
  \left(A_{j}{}^{ikr}\, A_{k}{}^{stu}
  +A^{ik}\delta_{j}^{r}A_{k}{}^{stu}-A^{ir}A_{j}{}^{stu}\right)  \;,
  \nonumber\\[1ex]
0&=& A^r{}_{ijk} \,A_{r}{}^{mnp} - 9\,A^{[m}{}_{r[ij}\,
A_{k]}{}^{np]r}
- 9\, \delta_{[i}{}^{[m} \, A^n{}_{|rs|j}\,A_{k]}{}^{p]rs} \nn\\
&&{}- 9\, \delta_{[ij}{}^{[mn}\,A^{|u|}{}_{k]rs}\, A_{u}{}^{p]rs}
+\delta{}_{ijk}{}^{mnp} \,A^u{}_{rst} \, A_{u}{}^{rst}  \;.
\label{quad}
\end{eqnarray}
Let us finally note that the scalar potential of the theory is given in terms of these tensors by
\bea
V&=& -\frac{3}{4}\,\left( A_{kl} A^{kl} - \frac{1}{18}\, A_{n}{}^{jkl} {A^n}_{jkl}\right)
\;,
\label{potential}
\eea
and that its extremal points are given by those values for the scalar fields
at which the tensor
\bea
{\cal C}_{ijkl}&=&
{A{}^m}_{[ijk} A_{l]m}
+ \frac34\, {A{}^m}_{n[ij} {A^n}_{kl]m}
\;,
\label{defC}
\eea
becomes anti-selfdual:
\bea
{\cal C}_{ijkl} + \frac1{24}\, \varepsilon_{ijklmnpq}\, {\cal C}^{mnpq}
&=& 0\;.
\label{ext}
\eea
At these extremal points, the couplings~(\ref{Yuk}) give rise to the
fermionic mass terms. For example,
the gravitino masses are obtained as the eigenvalues of
the properly normalized tensor $A_{ij}$.
For vanishing gauge fields and constant scalars, the Killing spinor equations
of the theory reduce to
\bea
0 &\stackrel{!}{\equiv}& \delta_{\epsilon} \psi_{\mu}^i ~=~
2\, {\cal D}_{\mu} \epsilon^i + \sqrt{2} \,A^{ij} \gamma_{\mu} \epsilon_j \;,\nonumber\\
0 &\stackrel{!}{\equiv}&\delta_{\epsilon} \chi^{ijk} ~=~
-2 \,A_l{}^{ijk} \epsilon^l \;.
\label{KSE}
\eea
Let us give, for the sake of  completeness, the mass matrices for the various fields \cite{LeDiffon:2011wt}.
The linearization of
the scalar field equations  yields, to lowest order,
\bea
\Box\, \delta\phi_{ijkl} &=&
{\cal M}_{ijkl}{}^{mnpq}\,\delta\phi_{mnpq} + {\cal O}(\delta\phi^2)
\;,
\eea
where $\delta\phi_{ijkl}$ are fluctuations  of the self-dual scalar fields $\phi_{ijkl}=\ft1{24}\varepsilon_{ijklpqrs}\, \phi^{pqrs}$ around their vacuum value $\phi_0=(\phi_0^{ijkl})$ and
the scalar mass matrix ${\cal M}_{ijkl}{}^{mnpq}$ is given by
\bea
 {\cal M}_{ijkl}{}^{mnpq}\,\delta\phi^{ijkl}\delta\phi_{mnpq} &=&
6\, \left(A_{m}{}^{ijk}A^{l}_{ijn}\!-\!\ft14A_{i}{}^{jkl}A^{i}_{jmn}\right)\delta\phi^{mnpq}\delta\phi_{klpq}
\nonumber\\
&&{}
+\left(\ft5{24}\,A_{i}{}^{jkl}A^{i}{}_{jkl}-\ft12A_{ij}A^{ij}\right) \delta\phi^{mnpq}\delta\phi_{mnpq}
\nonumber\\
&&{}
-\ft23\,A_{i}{}^{jkl} A^{m}{}_{npq} \, \delta\phi^{inpq}\delta\phi_{jklm}\nonumber\\[1ex]
&=& 12\,V^{(2)}(\delta\phi)
\;,
\label{Mscalar_sym}
\eea
where we have denoted by $V^{(2)}(\delta\phi)$ the terms on the scalar potential (\ref{potential})
which are second order in $\delta\phi$ upon expansion around the vacuum $\phi_0$:
\begin{equation}
V(\phi)~=~V_0+V^{(2)}(\delta\phi)+{\cal O}(\delta\phi^3)\,.
\end{equation}
The vector mass matrix reads
\bea
{\cal M}_{\rm vec} &=&
\left(
\begin{array}{cc}
{\cal M}_{ij}{}^{kl} & {\cal M}_{ijkl}
\\
 {\cal M}^{ijkl}&
 {\cal M}^{ij}{}_{kl}
\end{array}
\right)
\;,
\label{M_vector}
\eea
with
\bea
{\cal M}_{ij}{}^{kl}
&=&
-\ft16\,A_{[i}{}^{npq} \delta_{j]}^{[k} A^{l]}{}_{npq}
+\ft12\, A_{[i}{}^{pq[k}A^{l]}{}_{j]pq}\,,
\nonumber\\[2ex]
{\cal M}_{ijkl} &=&
\ft1{36}\,A_{[i}{}^{pqr} \epsilon_{j]pqrmns[k} A_{l]}{}^{mns}\,.
\eea
We can also give this matrix a manifestly symplectic covariant form
\begin{equation}
{\cal M}_{{\rm vec}\,M}{}^N=\frac{1}{6}\,\left[{\rm Tr}(X_M\,X_P)+{\rm Tr}(\mathbb{M}^{-1}\,X_M\,\mathbb{M}\,(X_P)^T)\right]\,\mathbb{M}^{PN}\,,
\end{equation}
where $\mathbb{M}_{MN}=(\mathcal{V}\,\mathcal{V}^T)_{MN}$ is the symmetric, symplectic, positive definite matrix constructed from the coset representative $\mathcal{V}_M{}^N$ in the ${\bf 56}$ of ${\rm E}_{7(7)}$. By virtue of the quadratic constraint (\ref{Xalgebra}) on $\Theta$, the matrix ${\cal M}_{\rm vec}$ always has 28 vanishing eigenvalues (corresponding to the magnetic vector fields), while the remaining eigenvalues define the masses of the (electric) vector fields.\par
Finally, the gravitino and fermion mass matrices are:
\begin{eqnarray}
{\cal M}_\psi{}^{ij} &=& \sqrt{2}\,A^{ij}
\;,\qquad
{\cal M}_\chi{}^{ijk,lmn} ~=~
\ft1{12} \sqrt{2}\, 
\epsilon^{ijkpqr[lm}  A^{n]}{}_{pqr}
\;.
\label{Mferm}
\end{eqnarray}
The first matrix ${\cal M}_\psi$ carries the information
about the breaking of supersymmetry and the
latter matrix has to be evaluated after projecting out the fermions
that are eaten by the massive gravitinos. Explicitly, at an AdS vacuum
and in a basis in which $A_{ij}$ is diagonal, the effective fermion mass matrix is given by
\bea
{\cal M}_\chi{}^{ijk,lmn} ~=~
\frac1{12} \sqrt{2}\, \Big(
\epsilon^{ijkpqr[lm}  A^{n]}{}_{pqr}~+~
 \frac{4}{3}\, \sum_{p,q}{}^\prime  A_p{}^{ijk} A_q{}^{lmn} \Big(\frac{A}{A^2 + {\bf 1}\, V/6}\Big)^{pq} \Big)
 \;,
\eea
with the sum running only over the massive gravitino directions.

\subsection{${\cal N}>2$ AdS vacua}
\label{subsec:N2}

We have reviewed, how a given embedding tensor defines the scalar potential (\ref{potential})
of gauged supergravity which in turn may carry extremal points (\ref{ext}) at which supersymmetry is
(partially) broken.
The embedding tensor formalism allows to nicely invert the problem and to search for vacua with
given properties by simultaneously scanning the set of all possible gaugings.
That strategy has e.g.\ been applied in \cite{Dibitetto:2011gm,DallAgata:2011aa,Borghese:2012qm,Borghese:2012zs,Borghese:2013dja}
in order to identify and analyze vacua with a given residual symmetry group.
Concretely, any joint solution to the quadratic equations (\ref{quad}) and the vacuum condition (\ref{ext}) defines a vacuum
in some maximal gauged supergravity. The associated embedding tensor and gauge group generators can then be
restored via (\ref{TAB}), (\ref{TT}), and (\ref{XMNK}).

In this section, we will investigate AdS vacua in maximal supergravity that preserve more than 2 supersymmetries.
Let us assume that the matrices $A_{ij}$, $A_i{}^{jkl}$ describe an AdS vacuum preserving ${\cal N}$ supersymmetries,
i.e.\ assume the existence of ${\cal N}$ independent solutions of (\ref{KSE}). Without loss of generality, we may then choose a basis
$i=(\alpha, a)$ in which
\bea
|A_{\alpha\beta}|&=&g\,\delta_{\alpha\beta}\;,\quad A_{\alpha a}~=~0 \;,\quad
A{}^\alpha{}_{ijk}~=~0\;,
\nonumber\\
&&{}
\mbox{for}\quad
\alpha,\beta = 1, \dots, {\cal N}\;,\quad
a={\cal N}+1, \dots, 8
\;,
\label{ansatz_susy}
\eea
and try to solve the quadratic equations (\ref{quad}), (\ref{ext}) under these assumptions for the remaining components of the
tensors $A_{ij}$, $A_i{}^{jkl}$\,.
First, let us note that for all non-vanishing ${\cal N}>0$, equations (\ref{ext}) follow directly as upon reduction of equations
(\ref{quad}) by (\ref{ansatz_susy}).
This is nothing but a remnant of the fact that the existence of a Killing spinor in general implies part of the remaining bosonic equations
of motion (in this case the scalar field equations for constant scalars).
It thus remains to solve equations (\ref{quad}) with the ansatz (\ref{ansatz_susy}). Since they are homogeneous, we may furthermore set $g=1$\,.
Some contraction of the first equation from (\ref{quad}) then allows to deduce the value of the potential as
\bea
V &=& -6\;.
\eea
On the other hand, the first equation of (\ref{quad}) with $k=\alpha, l=\beta, n=\gamma$ yields
\bea
 0&=&
  -2A{}^{m}{}_{\beta \gamma  \delta}\,A^{\alpha \delta} \quad
  \Longrightarrow\quad
  A{}^{m}{}_{\alpha\beta\gamma}~=~0
 \;,
 \label{null3}
\eea
thus imposes the absence of the components $A{}^{m}{}_{\alpha\beta\gamma}$.
For later use, we also note that
 the second equation of (\ref{quad}) in particular implies that
\bea
0 &=&
3\, A{}^a{}_{\alpha e[b} A{}^e{}_{cd]\beta} + A{}^a{}_{e\alpha\beta} A{}^e{}_{bcd}+
A_{\alpha\beta}\,A{}^a{}_{bcd}-\frac16\,\epsilon_{\alpha\beta \,bcd \, ijk} A^{ae} A_{e}{}^{ijk}
\;.
\label{q2l}
\eea

Let us now specialize to the case of ${\cal N}>2$ preserved supersymmetries. In this case,
the preserved supercharges transform in the vector representation of the AdS R-symmetry ${\rm SO}({\cal N})$.
We can then give a systematic discussion of these vacua according to the transformation of the
broken supercharges (i.e.\ the massive gravitino fields) under that ${\rm SO}({\cal N})$. In particular, all non-vanishing
components of the tensors $A_{ij}$, $A_i{}^{jkl}$ must be singlets under ${\rm SO}({\cal N})$.
Let us consider as an example the case when all broken supercharges are singlet under ${\rm SO}({\cal N})$.
If ${\cal N}>4$ this is the only option (in the absence of non-trivial ${\rm SO}({\cal N})$ representations of sufficiently small size).
The non-vanishing components of the tensors $A_{ij}$, $A_i{}^{jkl}$ are thus given by
\bea
\left\{ A_{\alpha\beta},\, A_{ab},\,A{}^a{}_{bcd} \right\}
\;,
\eea
with all other possible singlets under ${\rm SO}({\cal N})$ vanishing, in view of (\ref{ansatz_susy}) and (\ref{null3}).
Now (\ref{q2l}) immediately implies that also $A{}^a{}_{bcd}=0$, i.e.\ the entire tensor $A_i{}^{jkl}$ vanishes.
Then however the first equation of (\ref{quad}) implies that
\bea
A_{ac}\,A^{cb} &=& \delta_a{}^b
\;,
\eea
i.e.\ after diagonalisation the eigenvalues of $A_{ab}$ are of absolute value 1 and all correspond
to unbroken supersymmetries. The resulting vacuum thus is an ${\cal N}=8$ vacuum.
We conclude that there are no ${\cal N}>2$ AdS vacua (other than the ${\cal N}=8$ ones) if the
broken supercharges transform as singlets under the ${\rm SO}({\cal N})$ R-symmetry. In particular,
there are no AdS vacua in maximal supergravity preserving $4<{\cal N}<8$ supersymmetries. For $\mathcal{N}=6$ AdS vacua, this is consistent  with the result of \cite{Andrianopoli:2008ea}.

In the following, we will thus assume that the broken supercharges transform non-trivially under the
${\rm SO}({\cal N})$ R-symmetry and determine the general solution of (\ref{quad}) for ${\cal N}>2$.

\section{AdS ${\cal N}=8 \,\longrightarrow\, {\cal N}=3$  supersymmetry breaking patterns}\label{N83}
\label{sec:N3patterns}

\begin{table}[bt]
\begin{center}
{\small
\begin{tabular}{|c||c|c|c|c|}
\hline
$\Delta$  $\Big\backslash$ $s$   & $\frac32$ & $1$  & $\frac12$  & $0$ \\ \hline\hline
$E_0+3$ &&&&$[j]$ \\ \hline
$E_0+\frac52$ &&&$[j+1]+[j]\,$\textcolor{blue}{$+\,[j-1]$}&\\ \hline
$E_0+2$ &&
$[j+1]$\textcolor{blue}{$\,+\,[j]+[j-1]$}&&
\begin{tabular}{c}
$[j+2]+[j+1]+[j]$\\
\textcolor{blue}{${}+[j]+[j-1]+[j-2]$}
\end{tabular}
\\ \hline
$E_0+\frac32$ &\textcolor{blue}{$[j]$}&&
\begin{tabular}{c}
$[j+2]+[j+1]$\textcolor{blue}{${}+[j+1]$}\\
\textcolor{blue}{${}+2[j]+2[j-1]+[j-2]$}
\end{tabular}&\\ \hline
$E_0+1$ &&\textcolor{blue}{$[j+1]+[j]+[j-1]$}&&
\begin{tabular}{c}
$[j+2]$\textcolor{blue}{$\,+\,[j+1]+2[j]$}\\
\textcolor{blue}{${}+[j-1]+[j-2]$}
\end{tabular}\\ \hline
$E_0+\frac12$ &&&\textcolor{blue}{$[j+1]+[j]+[j-1]$}&\\ \hline
$E_0$ &&&&\textcolor{blue}{$[j]$}\\ \hline
\end{tabular}
}
\caption{\small The long ${\cal N}=3$ gravitino multiplet $DS(3/2,E_0,j)_{\rm L}$,
organized by energy~$\Delta$ and spin~$s$.
When the energy saturates the unitarity bound $E_0=j+1$,
the blue states in the table form a semi-short multiplet \textcolor{blue}{$DS(3/2,j+1,j)_{\rm S}$}
and the other states decouple as a vector multiplet $DS(1,j+2)$.}
\label{tab:Gmult}
\end{center}
\end{table}

Before we start the analysis of the ${\cal N}=3$ solutions of the quadratic constraints (\ref{quad}),
it is instructive to study the possible decompositions of the ${\cal N}=8$ supergravity multiplet
into ${\cal N}=3$ multiplets, i.e.\ to identify the possible kinetic scenarios of supersymmetry breaking.
The multiplet structure of the  ${\cal N}=3$ AdS supergroup $OSp(3|4)$ is well known,
see \cite{Freedman:1983na,Termonia:1999cs,Fre:1999xp}, in the following we will adopt the notation from \cite{Fre:1999xp}.
The relevant multiplets for our discussion are the massive gravitino multiplets which accommodate the five massive gravitinos after the
supersymmetry breaking ${\cal N}=8\longrightarrow {\cal N}=3$.
The structure of the generic long gravitino multiplet $DS(3/2,E_0,j)_{\rm L}$ is recollected in table~\ref{tab:Gmult}.
It is characterized by two numbers: the energy $E_0$ of its ground state, and the isospin $j$, characterising the
representation of the gravitino under the R-symmetry group ${\rm SO}(3)$. Unitarity imposes the bound $E_0\ge j+1$ for the
ground state energy (the ground state having spin 0). As usual for such supergroups, multiplet shortening occurs when the unitarity bound is saturated.
At this value of $E_0$, the multiplet splits into a short massive gravitino multiplet together with a vector multiplet according to
\bea
DS(3/2,E_0,j)_{\rm L}\Big|_{E_0 = j+1} &\longrightarrow&
DS(3/2,j+1,j)_{\rm S} + DS(1,j+2)
\;,
\eea
with the structure of the vector multiplet $DS(1,j+2)$ given in table~\ref{tab:Vmult}.
At low values of $j$, the multiplet structure becomes non-generic, but the tables still capture the correct
representation content upon formally extending the definition of ${\rm SO}(3)$ representations $[j]$ to negative $j$ according to\footnote{By $-[j-1]$ we mean that the isospin multiplet structure is obtained by deleting the representations with negative isospin ($[-j]$) and, for each of them, a representation $[j-1]$.}
\bea
[-j] &\equiv& -[j-1]\;.
\eea
In particular, the lowest-lying short gravitino multiplet $DS(3/2,1,0)_{\rm S}$
carries a massless gravitino, three massless vectors, three fermions, and two scalars of energy $1$ and $2$.
Due to the massless gauge fields, its presence in the spectrum implies an enhancement of supersymmetry
and gauge symmetry. Similarly, the massless vector multiplet $DS(1,1)$ carries six scalars together with
a massless vector and four fermions.

\begin{table}[bt]
\begin{center}
{\small
\begin{tabular}{|c||c|c|c|c|}
\hline
$\Delta$  $\Big\backslash$ $s$    & $1$  & $\frac12$  & $0$ \\ \hline\hline
$j_0+2$ &
&&
$[j_0-2]$
\\ \hline
$j_0+\frac32$ &&
$[j_0-1]+[j_0-2]$ & \\ \hline
$j_0+1$ &$[j_0-1]$ &&
$[j_0]+[j_0-1]+[j_0-2]$ \\ \hline
$j_0+\frac12$ &&
$[j_0]+[j_0-1]$ &\\ \hline
$j_0$ &&&$[j_0]$\\ \hline
\end{tabular}
}
\caption{\small The ${\cal N}=3$ vector multiplet $DS(1,j_0)$.}
\label{tab:Vmult}
\end{center}
\end{table}

\begin{table}[bt]
\begin{center}
{\small
\begin{tabular}{|c||c|c|c|c|}
\hline
$\Delta$  $\Big\backslash$ $s$    & $2^*$  & $\frac32^*$
& $1^*$  & $\frac12$   \\ \hline\hline
$3$ &[0]&&&
\\ \hline
$\frac52$ &&[1]&&
\\ \hline
$2$ &&&[1]&
\\ \hline
$\frac32$ &&&&[0]
\\ \hline
\end{tabular}
}
\caption{\small The ${\cal N}=3$ massless gravity multiplet $DS(2,3/2,0)_{\rm S}$.}
\label{tab:GN3}
\end{center}
\end{table}

With the multiplet structure given in tables~\ref{tab:Gmult}, \ref{tab:Vmult}, the possible supersymmetry
breaking patterns correspond to the different ways of splitting up the ${\cal N}=8$ supergravity multiplet into
${\cal N}=3$ multiplets. At this stage, we do not make any assumption about the energies of the various states
(other than those implied by unitarity). The ${\cal N}=8$ supergravity multiplet consists of the graviton, 8 gravitinos,
28 vectors, 56 fermions and 70 scalars. Upon subtracting the ${\cal N}=3$ supergravity multiplet $DS(2,3/2,0)_{\rm S}$,
given in table~\ref{tab:GN3},
we are left with 5 gravitinos, 25 vectors, 55 fermions and 70 scalars, to be packaged into ${\cal N}=3$ multiplets.
There are various options for the splitting of the five massive gravitinos into
${\rm SO}(3)$ R-symmetry representations:
\bea
{\rm I)} &5 ~\longrightarrow& 5\;,\nonumber\\
{\rm II)}&5 ~\longrightarrow& 3+1+1\;,\nonumber\\
{\rm III)}& 5 ~\longrightarrow& 2+2+1\;,
\label{var5}
\eea
where we have taken into account the reality property of the gravitinos which rules out decompositions such as $4+1$, $3+2$, etc..
Moreover, the general discussion of section~\ref{subsec:N2} has ruled out the trivial decomposition
$5\longrightarrow 1+1+1+1+1$\,.
Let us discuss the patterns (\ref{var5}) one by one.

Option I) in (\ref{var5}) leaves the five massive gravitinos in the irreducible spin-2 representation of ${\rm SO}(3)$.
According to table~\ref{tab:Gmult}, they can sit either in a long multiplet $DS(3/2,E_0,2)_{\rm L}$ or in its shortened version
$D(3/2,3,2)_{\rm S}$. Simple counting of states shows that the long multiplet carries 30 vector fields and thus does not fit into
${\cal N}=8$ supergravity. The short multiplet $DS(3/2,E_0,2)$ on the other hand does fit into ${\cal N}=8$ supergravity
with the remaining states filling precisely two vector multiplets.
A first possible kinetic pattern thus is given by
\bea
{\rm I)} &:& {\cal N}=8\;\longrightarrow\;
DS(2,3/2,0)_{\rm S} + DS(3/2,3,2)_{\rm S} + 2\cdot DS(1,1)
\;.
\label{optI}
\eea
The next option in  (\ref{var5}) is the partition $3+1+1$, in which case there are various possibilities depending on
the embedding of these gravitinos into the corresponding long or short gravitino multiplets.
Naive counting allows for the following possibilities
\bea
\begin{tabular}{c|cccc}
 II)&3&1&1&vectors\\\hline
 a):& L&S&S& 1\\
 b):& S&L&L& 0 \\
 c):& S&L&S& 3\\
 d):& S&S&S& 6 \\
\end{tabular}
\qquad\;.
\label{optII}
\eea
The last column denotes the number of vector multiplets that describe the remaining matter spectrum,
once the gravity and gravitino multiplets are subtracted from ${\cal N}=8$. Here, we note the following
property of the vector multiplet: when ignoring the energy of the states, the field content of $DS(1,j_0)$ coincides
with the tensor product of $DS(1,1)$ with the ${\rm SO}(3)$ representation $[j_0-1]$ of the vector fields.
As a consequence, for instance the 3 vectors in the third row of (\ref{optII}) can either correspond to three multiplets $DS(1,1)$
or to a single multiplet $DS(1,2)$, the field content only differs in energies.
Let us take a closer look at the decompositions of (\ref{optII}): the cases a) and d) both carry two gravitinos in the short massless
$DS(3/2,1,0)_{\rm S}$, i.e.\ both cases in fact correspond to a supersymmetry enhancement to ${\cal N}=5$.
Such vacua have been ruled out by the general discussion in section~\ref{sec:AdS} and cannot be dynamically realized.
We are thus left with the
options IIb) and IIc), of which the latter corresponds to a supersymmetry enhancement to ${\cal N}=4$.

The third option in  (\ref{var5}) is the partition $2+2+1$,
for which we find two possibilities
\bea
\begin{tabular}{c|cccc}
 III)&2&2&1&vectors\\\hline
 a):& S&S&L& 3\\
 b):& S&S&S& 6 \\
\end{tabular}
\qquad\;.
\label{optIII}
\eea

In summary, the possible AdS ${\cal N}=8 \,\longrightarrow\, {\cal N}=3$  supersymmetry breaking patterns
are given by the following decompositions of the ${\cal N}=8$ supergravity multiplet
\bea
{\rm I)} &:&
DS(2,3/2,0)_{\rm S} + DS(3/2,3,2)_{\rm S} + 2\cdot DS(1,1)
\;,
\label{opts}
\\
{\rm IIb)} &:&
DS(2,3/2,0)_{\rm S} + DS(3/2,2,1)_{\rm S} + DS(3/2,E_1,0)_{\rm L} + DS(3/2,E_2,0)_{\rm L}
\;,
\nonumber\\
{\rm IIc)} &:&
DS(2,3/2,0)_{\rm S} + DS(3/2,2,1)_{\rm S}+ DS(3/2,E_0,0)_{\rm L}+ DS(3/2,1,0)_{\rm S}  + DS(1,2)
\;,
\nonumber\\
{\rm IIIa)} &:&
DS(2,3/2,0)_{\rm S} + 2\cdot DS(3/2,3/2,1/2)_{\rm S}  + DS(3/2,E_0,0)_{\rm L} +  DS(1,2)
\;,
\nonumber\\
{\rm IIIb)} &:&
DS(2,3/2,0)_{\rm S} + 2\cdot DS(3/2,3/2,1/2)_{\rm S}  + DS(3/2,1,0)_{\rm S} +  2\cdot DS(1,2)
\;.
\nonumber
\eea
In the following we will study which of these patterns can actually be dynamically realized in ${\cal N}=8$
supergravity and determine the specific gaugings which allow for the corresponding vacua.

\section{${\cal N}=3$ and ${\cal N}=4$ vacua}
\label{sec:vacua}

\subsection{Solutions of the quadratic equations}
\label{solutions}

The ${\rm SO}(8)$ subgroup of ${\rm SU}(8)$ naturally splits into ${\rm SO}(3)\times {\rm SO}(5)$.
We  require the vacuum at the origin (and thus the  tensors $A_{ij},\,A^i{}_{jkl}$) to  be invariant under the diagonal group ${\rm SO}(3)_{\rm d}$ of the ${\rm SO}(3)$ group acting only on the Killing spinors and a second ${\rm SO}(3)$ embedded inside ${\rm SO}(5)$ according to the transformation of the massive gravitinos. We shall separately discuss the three cases corresponding to the allowed inequivalent embeddings (\ref{var5})
of ${\rm SO}(3)$ inside ${\rm SO}(5)$ and in each if them study the solutions to the system
(\ref{quad}). In all cases we have reduced the system by implementing the most general ansatz in terms of singlets under ${\rm SO}(3)_{\rm d}$ and (with the help of mathematica) systematically scanned the remaining equations for their real solutions. Such solutions turn out to be extremely rare.
Some computational details are relegated to appendix~\ref{app:details}.

\subsubsection{Case ${\bf 5}\rightarrow {\bf 5}$}
With this decomposition, there are six ${\rm SO}(3)_{\rm d}$ singlets in the tensors $A_{ij}$, $A_i{}^{jkl}$, three of which are
killed by the general discussion of section~\ref{subsec:N2}. It is straightforward to verify that the remaining system of quadratic equations
for three parameters does not possess any real solution (other than the known ${\cal N}=8$ solution), such that this possibility is ruled out by direct computation.

\subsubsection{Case ${\bf 5}\rightarrow {\bf 2}+{\bf 2}+{\bf 1}$}
Let us first split the $A,B,\dots$ indices into $\Lambda,\Sigma,\dots{}=4,5,6,7$ labeling the fundamental of the ${\rm SO}(4)$ inside ${\rm SO}(5)$, and identify the singlet in the decomposition with the value $i=8$. The index $i$ thus splits in $i=\alpha,\Lambda,8$. Next we embed ${\rm SO}(3)$ inside ${\rm SO}(4)$ by identifying its generators with the anti-self-dual matrices $(t^{(-)}_\alpha)^\Lambda{}_\Sigma$:
\begin{align}
t^{(-)}_1&=\left(
\begin{array}{llll}
 0 & \frac{1}{2} & 0 & 0 \\
 -\frac{1}{2} & 0 & 0 & 0 \\
 0 & 0 & 0 & -\frac{1}{2} \\
 0 & 0 & \frac{1}{2} & 0
\end{array}
\right)\,\,;\,\,\,t^{(-)}_2=\left(
\begin{array}{llll}
 0 & 0 & -\frac{1}{2} & 0 \\
 0 & 0 & 0 & -\frac{1}{2} \\
 \frac{1}{2} & 0 & 0 & 0 \\
 0 & \frac{1}{2} & 0 & 0
\end{array}
\right)\,\,;\,\,\,t^{(-)}_3=\left(
\begin{array}{llll}
 0 & 0 & 0 & -\frac{1}{2} \\
 0 & 0 & \frac{1}{2} & 0 \\
 0 & -\frac{1}{2} & 0 & 0 \\
 \frac{1}{2} & 0 & 0 & 0
\end{array}
\right)
\end{align}
We also have the complementary set of generators $({t}^{(+)}_{\hat{\alpha}})^\Lambda{}_\Sigma$, commuting with $t_\alpha$, obtained by changing the sign of the 4th row and columns of the latter. The following properties hold:
\begin{equation}
t^{(\pm)}_{\alpha \,\Lambda\Sigma}=\pm\frac{1}{2}\epsilon_{\Lambda\Sigma\Gamma\Delta}\,t^{(\pm)}_{\alpha \,\Gamma\Delta}\,.
\end{equation}
The ${\rm SO}(3)_{\rm d}$ generators in the ${\bf 8}$ of ${\rm SO}(8)$ read:
\begin{equation}
t_\alpha=\left(\begin{matrix}\epsilon_{\beta\alpha\gamma} & {\bf 0} & {\bf 0}\cr {\bf 0} & (t^{(-)}_{\alpha})_{\Lambda}{}^{\Sigma} & {\bf 0}\cr {\bf 0}& {\bf 0}& 0\end{matrix}\right)\,,
\end{equation}
and close the $\mathfrak{so}(3)_{\rm d}$ algebra:
\begin{equation}
[t_{\alpha},\,t_{\beta}]=\epsilon_{\alpha\beta\gamma}\,t_\gamma\,.\label{so3alg}
\end{equation}
With the general ansatz for $A_{ij}$, $A_i{}^{jkl}$ in terms of singlets under this ${\rm SO}(3)_{\rm d}$  (c.f.\ appendix~\ref{app:details}),
we find aside from the known ${\cal N}=8$ solution $A_i{}^{jkl}=0,\,A_{77}=\pm 1,\,\,A_{88}=e^{i\varphi}$,
only the following ${\cal N}=3$ solution:
\begin{align}
A_{\alpha\beta}&=\delta_{\alpha\beta}\;,\quad A_{\Lambda\Sigma}=\frac{3}{2}\,\epsilon\,\delta_{\Lambda\Sigma}\;,\quad
A_{88}=-\sqrt{3}\,e^{3i\varphi}\;,\quad\nonumber\\
A^\Lambda{}_{\Sigma\alpha\beta}&=\epsilon_{\alpha\beta\gamma}\,(t^{(-)}_{\gamma})^\Lambda{}_{\Sigma}\;,\quad
A^\Lambda{}_{\Sigma\alpha 8}=-\sqrt{3}\,e^{i\varphi}\,(t^{(-)}_{\alpha})^\Lambda{}_{\Sigma}\;,\quad,\nonumber\\
A^\Lambda{}_{\Sigma\Gamma\Delta}&=\epsilon \frac{\sqrt{3}}{2}\,e^{-i\varphi}\,\epsilon_{\Lambda\Sigma\Gamma\Delta}\;,\quad
A^8{}_{\alpha\beta\gamma}=0\;,\quad A^8{}_{8\Lambda\Sigma}=0\;,\quad A^8{}_{\alpha\Lambda\Sigma}=-2\,
\epsilon\,e^{-2i\varphi}\,\,t^{(-)}_{\alpha \,\Lambda\Sigma}\,,\label{s5221}
\end{align}
with real $\varphi$ and $\epsilon=\pm1$\,.
It effectively depends only  on the phase $\varphi$, since the sign $\epsilon=\pm 1 $ can be absorbed by an ${\rm SU}(8)$ transformation.
\subsubsection{Case ${\bf 5}\rightarrow {\bf 3}+{\bf 1}+{\bf 1}$}
Let us split the index $i$ into $i=\alpha,\alpha',a$, where $\alpha=1,2,3$, $\alpha'=4,5,6$ and $a=7,8$ is the index labeling the singlets.
The ${\rm SO}(3)_{\rm d}$ generators in the ${\bf 8}$ of ${\rm SO}(8)$ read:
\begin{equation}
t_\alpha=\left(\begin{matrix}\epsilon_{\beta\alpha\gamma} & {\bf 0} & {\bf 0}\cr {\bf 0} & \epsilon_{\beta'\alpha\gamma'} & {\bf 0}\cr {\bf 0}& {\bf 0}& {\bf 0}_2\end{matrix}\right)\,,
\end{equation}
and satisfy the relations (\ref{so3alg}).
With the general ansatz for $A_{ij}$, $A_i{}^{jkl}$ in terms of singlets under this ${\rm SO}(3)_{\rm d}$  (c.f.\ appendix~\ref{app:details}),
we find aside from the known ${\cal N}=8$ solution
only the following solution:
\begin{align}
A_{\alpha\beta}&=\delta_{\alpha\beta}\;,\quad A_{\alpha'\beta'}~=~2\xi\,\delta_{\alpha'\beta'} \;,\quad
A_{77}=2\,\eta\;,\quad A_{88}=\xi\,\eta\,e^{i\varphi}\,,\nonumber\\
A^7{}_{\alpha'\beta'\gamma}&=\sqrt{2}\,\xi\,\eta\,e^{-i\frac{\varphi}{4}}\,
\epsilon_{\alpha'-3\,\beta'-3\,\gamma}\;,\quad A^{\alpha'}{}_{\beta'\gamma 7}=-\sqrt{2}\,e^{-i\frac{\varphi}{4}}\,\epsilon_{\alpha'-3\,\beta'-3\,\gamma}\,,\nonumber\\
A^{\alpha'}{}_{\beta'\gamma' 8}&=\sqrt{2}\,\xi\,e^{i\frac{\varphi}{4}}\,\epsilon_{\alpha'-3\,\beta'-3\,\gamma'-3}\;,\quad
A^{\alpha'}{}_{\alpha ab}=-\eta\,e^{i\frac{\varphi}{2}}\,\delta^{\alpha'-3}_\alpha\,\epsilon_{ab}\,,\nonumber\\
A^7{}_{8\alpha'\alpha}&=-\xi\,e^{i\frac{\varphi}{2}}\,\delta_{\alpha'-3\,\alpha}\;,\quad
A^{\alpha'}{}_{\beta' \alpha\beta}=-2\,\delta^{\alpha'-3\,\beta'-3 }_{\alpha\beta}\,,\label{s53111}
\end{align}
where $\eta,\,\xi=\pm 1$. The parameter $\xi$ can be disposed of by means of a ${\rm SU}(8)$ transformation while the sign $\eta$ can be changed by shifting $\varphi\rightarrow \varphi+2\pi$. We can thus set $\xi=\eta=+1$.
Notice that $A^8{}_{ijk}=0$ which implies that this is actually an ${\cal N}=4$ solution and that the residual symmetry group is enhanced to ${\rm SO}(4)$.

\subsection{Gauge Groups and ${\rm E}_{7(7)}$-Invariants}

We have identified two AdS vacua in maximal supergravity by solving the system of quadratic constraints (\ref{quad}) for the
embedding tensor.
As the next step, we will have to determine the associated gauge groups, i.e.\ identify in which gauged maximal supergravity these vacua live.
We can compute the associated gauge group generators  via (\ref{TAB}), (\ref{TT}), and (\ref{XMNK}).
Much of the structure of the gauge group can already be inferred from
the ${\rm E}_{7(7)}$-invariant signature of the (generalized) Cartan-Killing metric
\begin{equation}
\mbox{sign}[{\rm Tr}(X_M\cdot X_N)]\,.
\label{CK}
\end{equation}
The above matrix has $28$ vanishing eigenvalues (due to the locality constraint (\ref{Xalgebra})) while the other $28$ eigenvalues define the  Cartan-Killing metric of the gauge algebra.

\subsubsection{The ${\cal N}=4$ vacuum}

We first compute the Cartan-Killing metric (\ref{CK}) for the ${\cal N}=4$ vacuum (\ref{s53111})
as a function of the angular parameter $\varphi$.
This allows the following identification of the corresponding underlying gauge group:
\bea
\begin{tabular}{|c|c|c|}
  \hline
 parameter & signature of C.-K. metric & gauge group\\\hline
 $\varphi=2\pi $& $(1_+,15_-,12_0)$ & $[{\rm SO}(1,1)\times {\rm SO}(6)]\ltimes T^{12}$ \\
  $0\le \varphi< 2\pi$&$(7_+,21_-)$ & ${\rm SO}(1,7)$ \\
  \hline
\end{tabular}
\label{tab4}
\eea
where $T^{12}$ denotes a subgroup generated by twelve nilpotent operators.
 Notice that AdS vacua in theories with gauged ${\rm SO}(1,7)$ and $[{\rm SO}(1,1)\times {\rm SO}(6)]\ltimes T^{12}$ groups were found in \cite{DallAgata:2011aa}. The residual supersymmetry, symmetry group (${\rm SO}(4)$) and spectrum of our vacuum distinguishes it from those found in the same reference.

 The gauge group ${\rm SO}(1,7)$ alone is not sufficient to determine the gauged supergravity, since there
 is a one-parameter class of such theories~\cite{DallAgata:2012bb,DallAgata:2014ita}. Rather, we expect to find a mapping between the
 angular parameter $\varphi$ which defines our vacua, and the $\omega$-angle that labels the one-parameter class of ${\rm SO}(1,7)$ theories.
 To this end, we compute other  ${\rm E}_{7(7)}$-invariants on the ${\cal N}=4$ vacuum, to compare with the same quantities evaluated for the $\omega$-rotated ${\rm SO}(1,7)$ gauge group.
 In particular, we consider the $1540\times 1540$ matrix:
 \begin{equation}
 \mathbb{K}_{MN}{}^{PQ}\equiv \frac{1}{4}\,d^{R_1 R_2 R_3 R_4}\,X_{R_1 M}{}^K\,X_{R_2NK}\,X_{R_3}{}^{PL}\,X_{R_4 L}{}^Q\,,
 \label{K}
 \end{equation}
 quartic in the gauge group generators (\ref{XMNK}),
which is antisymmetric in $[MN]$ and $[PQ]$, by virtue of the total symmetry of the ${\rm E}_{7(7)}$-invariant tensor $d^{R_1 R_2 R_3 R_4}$. This tensor $\mathbb{K}$ is related  to one computed in \cite{DallAgata:2012bb} for the $\omega$-deformed ${\rm SO}(8)$ gauging.
 Instead of evaluating the eigenvalues of this matrix, as was done for the corresponding tensor in \cite{DallAgata:2012bb} for a specific gauging, we evaluate the traces of its powers. We compute them for our ${\cal N}=4$ solution and for the ($\omega$-deformed) ${\rm SO}(1,7)$ and ${\rm SO}(8)$ gauging.

 For the invariant $d$-tensor  we use the following form in the ${\rm SU}(8)$-basis:
\begin{align}
d^{MNPQ}\,\Lambda_M\Lambda_N\Lambda_P \Lambda_Q&=\Lambda^{i_1 i_2}\Lambda^{i_3 i_4}\Lambda^{i_5 i_6}\Lambda^{i_7 i_8}\,\epsilon_{i_1\dots i_8}+\Lambda_{i_1 i_2}\Lambda_{i_3 i_4}\Lambda_{i_5 i_6}\Lambda_{i_7 i_8}\,\epsilon^{i_1\dots i_8}+\nonumber\\
&+96\,{\rm Tr}(\Lambda\bar{\Lambda}\Lambda\bar{\Lambda})-24\,{\rm Tr}(\Lambda\bar{\Lambda})^2\,,\nonumber\\
\Lambda_M& \equiv(\Lambda,\bar{\Lambda})  ~\equiv~ (\Lambda_{ij},\Lambda^{ij})\;.
\end{align}
For the ${\rm SO}(8)$ and ${\rm SO}(1,7)$ gaugings
the $X$-tensor (\ref{XMNK})  is computed via (\ref{TT}), (\ref{TAB}),  starting from  fermion shift tensors of the form
\cite{DallAgata:2011aa}:
\begin{equation}
A_{ij}=e^{i\,\omega}\,{\rm Tr}(\theta)\,\delta_{ij}\,\,\,,\,\,\,\,A_{i}{}^{jkl}=e^{-i\,\omega}\,(\Gamma_i{}^{jkl})_{IJ}\theta^{IJ}\,,
\end{equation}
where
\begin{equation}
\theta^{IJ}={\rm diag}(1,1,1,1,1,1,1,\kappa)\,,
\end{equation}
with $\kappa=+1$ for ${\rm SO}(8)$ and $-1$ for ${\rm SO}(1,7)$. We find for the traces of the various powers of (\ref{K})
\begin{align}
{\rm Tr}(\mathbb{K})&=0\,,\nonumber\\
{\rm Tr}(\mathbb{K}^2)&=2^{23}\times 3^4\times 5\times 7\times\left(7 (5 \kappa +3)+28 (\kappa -1) \cos (4 \omega )+(\kappa +7) \cos (8 \omega )\right)\,,\nonumber\\
{\rm Tr}(\mathbb{K}^3)&=2^{36}\times 3^7\times 5\times 7\times \left(35 \kappa +4 (7 \kappa +1) \cos (4 \omega )+(\kappa -1) \cos (8 \omega )-3\right) \sin ^2(2 \omega )\,,\nonumber\\
{\rm Tr}(\mathbb{K}^4)&\propto {\rm Tr}(\mathbb{K}^2)^2\,.\label{trKn}
\end{align}
Notice that in the ${\rm SO}(8)$ case these invariants have half-period $\pi/8$, namely they assume all possible values in the interval $\omega\in(0,\pi/8)$, while for the ${\rm SO}(1,7)$ gauging the half-period is $\pi/4$. In the former case we have independent gaugings only for $\omega\in(0,\pi/8)$, while in the latter case for $\omega\in(0,\pi/4)$, consistently with the results of \cite{DallAgata:2012bb,DallAgata:2012sx}.
Eq.s (\ref{trKn})  do not hold for $\kappa=0$, corresponding to ${\rm ISO}(7)$, in which case all traces are zero.\par
On our ${\cal N}=4$ solution (\ref{s53111}), (denoting the corresponding tensor by $\mathbb{K}_s$) these traces become
\begin{align}
{\rm Tr}(\mathbb{K}_s)&=0\,,\nonumber\\
{\rm Tr}(\mathbb{K}_s^2)&=2^{6}\times 3^4\times 5\times 7\times \left(3 \cos \left(\frac{\varphi }{2}\right)+\cos (\varphi )+2\right)\,,\nonumber\\
{\rm Tr}(\mathbb{K}_s^3)&=-2^{10}\times 3^7\times 5\times 7\times \cos ^4\left(\frac{\varphi }{4}\right)\,\nonumber\\
{\rm Tr}(\mathbb{K}_s^4)&\propto {\rm Tr}(\mathbb{K}_s^2)^2\,.
\label{tr4}
\end{align}
These expressions are symmetric under $\varphi\rightarrow -\varphi$ and $\varphi\rightarrow \varphi+4\pi$, so that they assume all possible values in the interval $(0,2\pi)$. Following the same reasoning as for the ${\rm SO}(1,7)$ and ${\rm SO}(8)$ cases, we can then argue that $X$ tensors in this class with generic $\varphi$ can be ${\rm SU}(8)$-rotated to one with $\varphi\in (0,2\pi)$.
In comparing the traces for the ${\cal N}=4$ vacuum (\ref{tr4}) to those of the ${\rm SO}(1,7)$ gauging (\ref{trKn}), we assume that
\begin{equation}
X^{(s)}_{MN}{}^P=\lambda(\varphi)\,{\rm E}_{7(7)}\star(X_{MN}{}^P)\,,\label{XsX}
\end{equation}
where $X^{(s)}_{MN}{}^P$ is the $X$-tensor on the ${\cal N}=4$ vacuum, ${\rm E}_{7(7)}\star(X_{MN}{}^P)$ is the ${\rm E}_{7(7)}$-rotated $X$ tensor of the ${\rm SO}(1,7)$ gauging, and we allowed for a proportionality factor $\lambda(\varphi)$ depending on the parameter $\phi$. Clearly the traces of $\mathbb{K}$ do not depend on the ${\rm E}_{7(7)}$-rotation, so that eq. (\ref{XsX}) implies:
\begin{align}
{\rm Tr}(\mathbb{K}_s^2)&=\lambda(\varphi)^8\,{\rm Tr}(\mathbb{K}^2)\,\,\,\,,\,\,\,\,\,{\rm Tr}(\mathbb{K}_s^3)=\lambda(\varphi)^{12}\,{\rm Tr}(\mathbb{K}^3)\,.\label{KsK}
\end{align}
We immediately realize that for $\omega=0$ the above system has no  solution since ${\rm Tr}(\mathbb{K}_s^2)=0$ while ${\rm Tr}(\mathbb{K}^2)\neq 0$, thus implying $\lambda=0$. Similarly for $\varphi=2\pi$, the second of (\ref{KsK}) implies $\omega=0$ while in the first ${\rm Tr}(\mathbb{K}_s^2)=0$ while ${\rm Tr}(\mathbb{K}^2)\neq 0$, againg implying $\lambda=0$.
This is compatible with our finding (\ref{tab4}) that the gauge group at $\varphi=2\pi$ degenerates to $[{\rm SO}(1,1)\times {\rm SO}(6)]\ltimes T^{12}$.

\begin{figure}[ht!]
 \centering
\includegraphics[scale=.6]{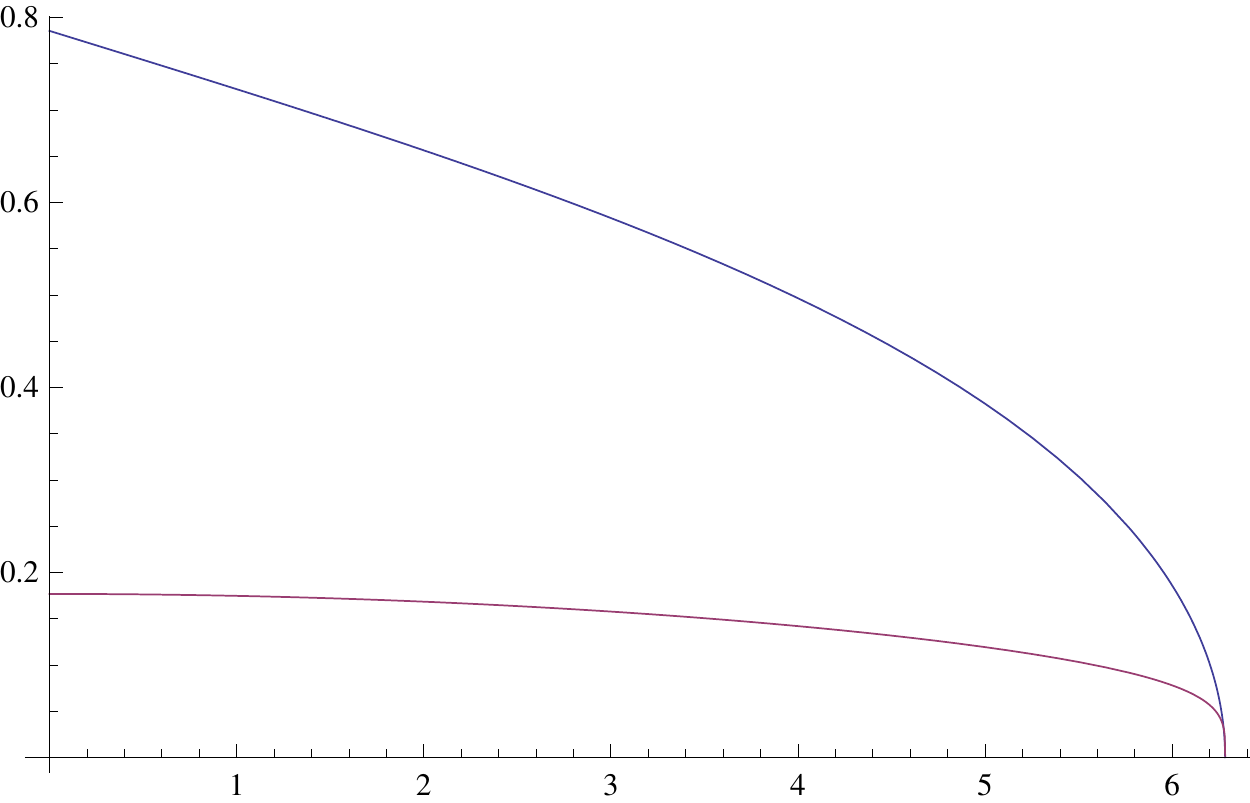}
\caption{{\small The parameters $\omega$ (blue) and $\lambda$ (red) as function of $\varphi$ for
the ${\cal N}=4$ vacuum.
The gauge group is ${\rm SO}(1,7)$ except for the point $\varphi=2\pi$ where $\lambda$ vanishes and the gauge group
degenerates to $[{\rm SO}(1,1)\times {\rm SO}(6)]\ltimes T^{12}$.
}}
\label{N4}
\end{figure}

For generic values of $\omega$ and $\varphi$ the relation between the two parameters can then be deduced from the ($\lambda$-independent) equation
\begin{equation}
\frac{{\rm Tr}(\mathbb{K}_s^3)^2}{{\rm Tr}(\mathbb{K}_s^2)^3}=\frac{{\rm Tr}(\mathbb{K}^3)^2}{{\rm Tr}(\mathbb{K}^2)^3}\,\,\Longleftrightarrow\,\,\,\,\frac{9 \cos ^2\left(\frac{\varphi }{4}\right)}{70 \left(2 \cos \left(\frac{\varphi
   }{2}\right)+1\right)^3}=\frac{144\, (\cos (4 \omega )+3)^4 \sin ^4(2 \omega )}{35\, (-28 \cos (4 \omega )+3 \cos (8 \omega
   )-7)^3}\;.
   \label{num4}
\end{equation}
whose solution $\omega(\varphi)$ is plotted in Fig.~\ref{N4}.
We conclude that the ${\cal N}=4$ vacuum can be found (for generic values of $\varphi$), only in the $\omega$-rotated ${\rm SO}(1,7)$ gauging, while in the limit $\omega\rightarrow 0$ it disappears. At the corresponding point ${\varphi=2\pi}$ in the parameter space of solutions it turns into a vacuum of the gauging with non-semisimple gauge group $[{\rm SO}(1,1)\times {\rm SO}(6)]\ltimes T^{12}$.

\subsubsection{The ${\cal N}=3$ vacuum}

\begin{figure}[ht!]
 \centering
\includegraphics[scale=.6]{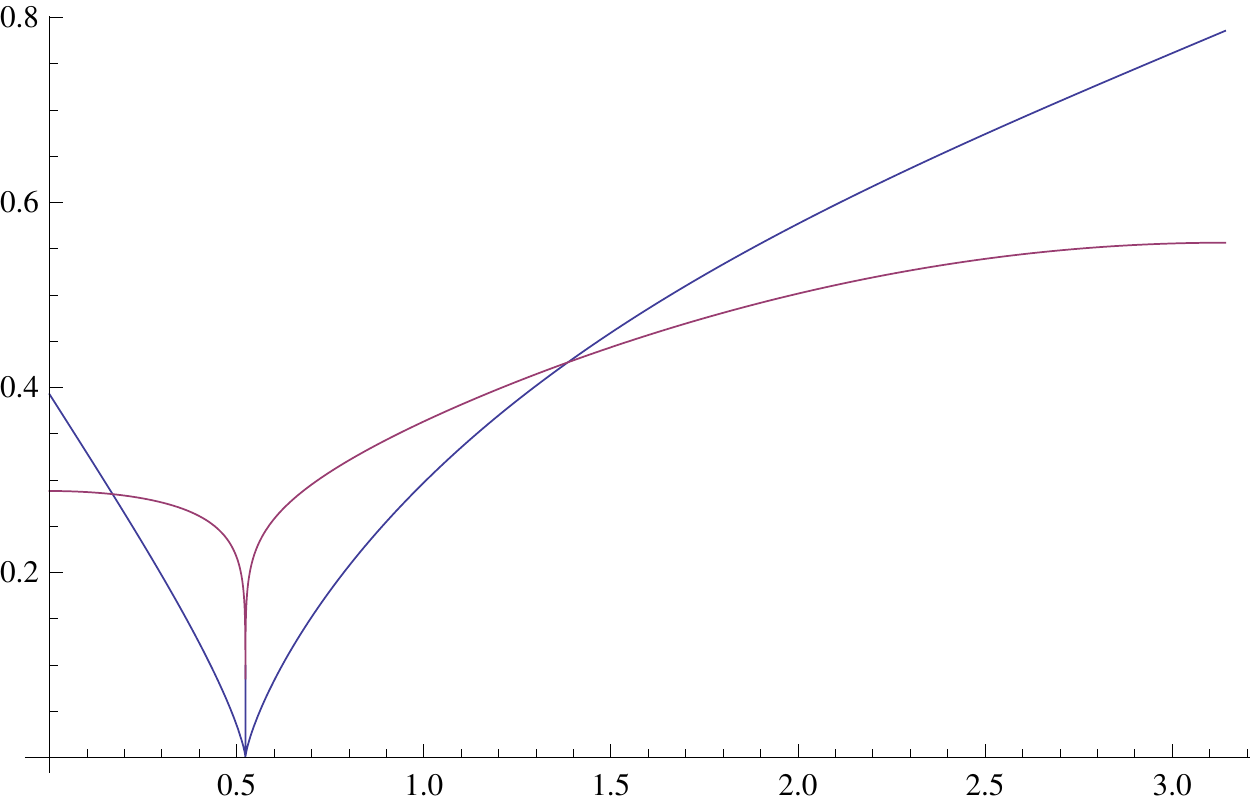}
\caption{{\small The parameters $\omega$ (blue) and $\lambda$ (red) as function of $\varphi$ for
the ${\cal N}=3$ vacuum.
The gauge group is ${\rm SO}(8)$ for $\varphi<\pi/6$ and ${\rm SO}(1,7)$ for $\varphi>\pi/6$. A the point $\varphi=\pi/6$
where $\lambda$ vanishes, the gauge group degenerates to ${\rm ISO}(7)$.
}}
\label{N3}
\end{figure}

The same analysis can be repeated for the ${\cal N}=3$ vacuum (\ref{s5221}).
In this case the computation of the Cartan-Killing metric (\ref{CK}) for the gauge group
indicates the following correspondence  between the values of $\varphi$ and the gauge group
\bea
\begin{tabular}{|c|c|c|}
  \hline
 parameter & signature of C.-K. metric & gauge group\\\hline
 $0\le \varphi<\frac{\pi}{6} $& $(0_+,28_-)$ & ${\rm SO}(8)$ \\
  $\frac{\pi}{6} < \varphi\le\pi $&$(7_+,21_-)$ & ${\rm SO}(1,7)$ \\
   $ \varphi=\frac{\pi}{6}$&$(0_+,21_-, 7_0)$ & ${\rm ISO}(7)$ \\
  \hline
\end{tabular}
\label{tab3}
\eea
Computing the traces of the tensor (\ref{K}) on
our ${\cal N}=3$ solution we find in this case:
\begin{align}
{\rm Tr}(\mathbb{K}_s)&=0\,,\nonumber\\
{\rm Tr}(\mathbb{K}_s^2)&=-2^{-1}\times 3^8\times 5\times 7\times \cos (\varphi ) \left(\sqrt{3} (\cos (2 \varphi )+3)-7 \cos (\varphi )\right)\,,\nonumber\\
{\rm Tr}(\mathbb{K}_s^3)&=2^{-2}\times 3^{11}\times 5\times 7\times (24 \sqrt{3} \cos (\varphi )-18 \cos (2 \varphi )+2 \sqrt{3} \cos (3 \varphi )-27)\,,\nonumber\\
{\rm Tr}(\mathbb{K}_s^4)&\propto {\rm Tr}(\mathbb{K}_s^2)^2\,,
\label{tr3}
\end{align}
which should now be compared to (\ref{trKn}) for $\kappa=\pm1$ in the different intervals of (\ref{tab3}).
The expressions of (\ref{tr3}) are symmetric under $\varphi\rightarrow -\varphi$ and $\varphi\rightarrow \varphi+2\pi$, so that they assume all possible values in the interval $(0,\pi)$. We then argue that an $X$-tensor in this class with a generic $\varphi$ can be ${\rm SU}(8)$-rotated to one within $\varphi\in (0,\pi)$.
The correspondence between $\varphi$ and $\omega$ for the $\omega$-rotated ${\rm SO}(1,7)$ and ${\rm SO}(8)$ groups is
obtained from an equation analogous to (\ref{num4}) and plotted in Fig. \ref{N3}. This illustrates that $\lambda$ vanishes, as $\omega\rightarrow 0$ ($\varphi\rightarrow \pi/6$), so that the vacuum disappears from the corresponding gauged theories. At this point in the $\varphi$ parameter space it becomes a vacuum of an ${\rm ISO}(7)$ gauged theory. Indeed, for $\varphi=\pi/6$, all traces of (\ref{tr3}) vanish.\\

 The existence  of the proportionality parameter $\lambda(\varphi)$ depending on $\varphi$ (or, equivalently on $\omega$) is due to the fact that we have fixed the value  the potential in our vacua to a given value ($-6$) by choosing the coupling constant. This function therefore encodes the dependence, in the corresponding $\omega$-rotated theories, of the cosmological constant of these vacua on $\omega$, which is a generic feature of all extrema of the potential aside from the $\mathcal{N}=8$ one \cite{DallAgata:2012bb,Borghese:2012zs}.

\subsection{Mass spectra}
We can eventually compute the spectra around the new vacua by evaluating the mass formulas (\ref{Mscalar_sym}),
(\ref{M_vector}), and (\ref{Mferm}) for our solutions $A_{ij}$, $A_i{}^{jkl}$ and
compare the result to the general multiplet structure discussed in section~\ref{sec:N3patterns}.
We find that in all cases the spectra are independent of the parameter $\varphi$.

\subsubsection{The ${\cal N}=3$ vacuum}
The scalar mass spectrum on the ${\cal N}=3$ vacuum is:
\begin{align}
m^2\,L_0^2\qquad&: 1\times\left(3(1+\sqrt{3})\right)\,;\,\,6\times\left(1+\sqrt{3}\right)\,;\,\,1\times\left(3(1-\sqrt{3})\right)\,;\,\,
6\times\left(1-\sqrt{3}\right)\,;\nonumber\\
&4\times\left(-\frac{9}{4}\right)\,;\,\,18\times\left(-2\right)\,;\,\,12\times\left(-\frac{5}{4}\right)\,;\,\,22\times\left(0\right)\,,
\end{align}
in units of the inverse anti- de Sitter radius $1/L_0$ from (\ref{L0}).
The Breitenlohner-Freedman bound $m^2\,L_0^2\ge -\frac{9}{4}$ \cite{Breitenlohner:1982jf} is satisfied by virtue of supersymmetry.
The normalized vector masses are given by:
\begin{align}
m^2\,L_0^2\quad&: 3\times\left(3+\sqrt{3}\right)\,;\,\,3\times\left(3-\sqrt{3}\right)\,;\,\,4\times\left(\frac{15}{4}\right)\,;\,\,
12\times\left(\frac{3}{4}\right)\,;\,\,6\times\left(0\right)\,,
\end{align}
The 22 massless scalar fields are the Goldstone bosons for the massive vector fields.
Together, we conclude that the ${\cal N}=3$ vacuum realizes option IIIa) from (\ref{opts}) with one long spin $3/2$ multiplet of
energy $E_0=\sqrt{3}$
\bea
DS(2,3/2,0)_{\rm S} + 2\cdot DS(3/2,3/2,1/2)_{\rm S}  + DS(3/2,\sqrt{3},0)_{\rm L} +  3\cdot DS(1,1)
\eea
The three extra massless vectors describe an extra ${\rm SO}(3)$ symmetry.
Explicit computation of the fermionic mass matrices (\ref{Mferm}) also confirms the multiplet structure.

\subsubsection{The ${\cal N}=4$ vacuum}
The scalar mass spectrum on the ${\cal N}=4$ vacuum is:
\begin{align}
m^2\,L_0^2&:1\times\left(10\right)\,;\,\,10\times\left(4\right)\,;\,\,11\times\left(-2\right)\,;\,\,48\times\left(0\right)\,.
\end{align}
The vector masses are
\begin{align}
m^2\,L_0^2&: 7\times\left(6\right)\,;\,\,15\times\left(2\right)\,;\,\,6\times\left(0\right)\,,
\end{align}
22 of the  massless scalar fields are the Goldstone bosons for the massive vector fields, while the six massless vectors gauge the residual ${\rm SO}(4)$ group.
This solution thus realizes option IIc) from (\ref{opts})
with a long spin $3/2$ multiplet of energy $E_0=2$
\bea
DS(2,3/2,0)_{\rm S}+ DS(3/2,1,0)_{\rm S} + DS(3/2,2,1)_{\rm S}+ DS(3/2,2,0)_{\rm L}  + DS(1,2)
\;,
\eea
and supersymmetry enhancement to ${\cal N}=4$,
under which the first two multiplets combine into the ${\cal N}=4$
massless supergravity multiplet and the remaining three multiplets
combine into a single ${\cal N}=4$ massive spin $3/2$ multiplet.
Again, an explicit computation of the fermionic mass matrices (\ref{Mferm})
confirms this multiplet structure.

\section{Conclusions}
In this paper we have studied AdS vacua of maximal supergravity in four dimensions with residual $\mathcal{N}>2$ supersymmetry. We exclude on general grounds $8>\mathcal{N}>4$  vacua and find two 1-parameter classes of $\mathcal{N}=3$ and $4$ vacua, which  can be embedded only in the $\omega$-rotated gauged models. Of particular importance are the models with ${\rm SO}(8)$ gauging  since they exhibit in addition an $\mathcal{N}=8$ vacuum. The eleven dimensional origin of the latter is still debated and in \cite{DallAgata:2012bb} it was conjectured to corrrespond to certain to ABJ theories \cite{Aharony:2008gk}, through  the AdS/CFT duality \cite{Maldacena:1997re}. Understanding the higher dimensional origin of the new $\mathcal{N}=3$ and $4$ vacua  is an important problem which deserves investigation.\par
Still in the light of the AdS/CFT correspondence, these new vacua should describe conformal fixed points of some dual (three-dimensional) field theory.  It would be also interesting, in this respect, to study RG flows between the conformal critical points dual to the two kinds of vacua in the $\omega$-deformed ${\rm SO}(1,7)$ models, or interpolating between the  $\mathcal{N}=8$ and $\mathcal{N}=3$ vacua in the $\omega$-deformed ${\rm SO}(8)$ theories, thus generalizing the analysis of \cite{Guarino:2013gsa,Tarrio:2013qga}.\par
An other issue which deserves investigation is the study of black holes asymptoting the new $\mathcal{N}=3$ and $4$ vacua, along the lines of \cite{Anabalon:2013eaa}.
It would also be interesting to understand to which extend the methods developed in this paper can be extended to a systematic analysis of the
AdS (and Minkowski) vacuum in maximal supergravity with ${\cal N}=2$ and ${\cal N}=1$ supersymmetry.

\section*{Acknowledgements}
We wish to thank D. Roest for helpful and inspiring discussions.
\bigskip
\bigskip
\bigskip

\appendix
\section{Normalizations and Conventions}
\label{app:not}
Let us recall the relevant notations for the scalar masses. The bosonic Lagrangian of $\mathcal{N}=8$ supergravity reads (setting $\kappa^2=8\pi G=1$)\footnote{We use the notations of \cite{deWit:2007mt}.}:
\begin{equation}
\mathcal{L}=e\,\left[-\frac{R}{2}+\frac{1}{12}\,P_\mu^{ijkl}\,P^{\mu}_{ijkl}+\frac{1}{4}\,I_{\Lambda\Sigma}(\phi)\,
F^\Lambda_{\mu\nu}F^{\Sigma\,\,\mu\nu}+\frac{1}{8\,e}\,\epsilon^{\mu\nu\rho\sigma}\,R_{\Lambda\Sigma}(\phi)\,
F^{\Lambda}_{\mu\nu}F^{\Sigma}_{\rho\sigma}-V(\phi)\right]\,.\label{bosl}
\end{equation}
We choose the vacuum at the origin $\phi_0=0$ in which the scalar potential is negative $V_0=V(0)<0$ (AdS vacuum), and  we expand about it:
\begin{equation}
\phi^{ijkl}=\phi_0^{ijkl}+\delta \phi^{ijkl}=\delta\phi^{ijkl}\,.
\end{equation}
Being interested in the scalar kinetic and mass terms, we set $A_\mu^\Lambda=0$ so that vielbein $P_\mu^{ijkl}$ on the vacuum reads:
\begin{equation}
P_\mu^{ijkl}=\partial_\mu\delta\phi^{ijkl}\,.
\end{equation}
Let us denote by $\phi^\alpha$, $\alpha=1,\dots, 70$ the real and imaginary parts of $\phi^{ijkl}$ subject to the self-duality condition. On the vacuum the kinetic and mass terms of the scalar fluctuations read:
\begin{equation}
\mathcal{L}_s^{(2)}=4\,\sum_\alpha\partial_\mu\delta \phi^\alpha\partial^\mu\delta \phi^\alpha-\frac{1}{2}\left.\frac{\partial^2 V}{\partial\phi^\alpha\partial\phi^\beta}\right\vert_{\phi=0}\,\delta \phi^\alpha\delta \phi^\beta\,,
\end{equation}
from which we deduce the mass matrix:
\begin{equation}
(m^2)_{\alpha\beta}=\frac{1}{8}\left.\frac{\partial^2 V}{\partial\phi^\alpha\partial\phi^\beta}\right\vert_{\phi=0}=\frac{1}{8} \,\frac{\partial^2 V^{(2)}}{\partial\delta\phi^\alpha\partial\delta\phi^\beta}\,,
\end{equation}
where $V^{(2)}(\delta\phi)$ is given in (\ref{Mscalar_sym}).\par
Four-dimensional anti-de Sitter space can be defined  as the connected hyperboloid in $\mathbb{R}^5$ described by the equation:
\begin{equation}
\eta_{AB}y^A\,y^B=L_0^2\,\,;\,\,\,\eta=\mbox{diag}(+---+)\,,
\end{equation}
$L_0$ being the ``radius'' of the AdS space-time. The Ricci tensor reads:
\begin{equation}
R_{\mu\nu}=-\Lambda\,g_{\mu\nu}\,\,,\,\,\,\,\,\Lambda=-\frac{3}{L_0^2}<0\,,
\label{L0}
\end{equation}
where $\Lambda$ is the cosmological constant. From (\ref{bosl}) we can identify:
\begin{equation}
\Lambda=V_0=-\frac{3}{L_0^2}\,.
\end{equation}
If $m^2$ is a generic eigenvalue of $(m^2)_{\alpha\beta}$, stability of the vacuum implies the following condition:
\begin{equation}
m^2\,L_0^2=\frac{3}{|V_0|}\,m^2\ge -\frac{9}{4}\,,
\end{equation}
which is the Breitenlohner-Freedman bound \cite{Breitenlohner:1982jf}.\par
For the reader's convenience we give the relations between the AdS energy $E_0$ and the masses of the various fields \cite{Freedman:1999gp}:
\begin{align}
\mbox{scalars}&:\,\,E_0=\frac{1}{2}\,\left(3\pm \sqrt{9+4\,m^2\,L_0^2}\right)\,,\nonumber\\
\mbox{vectors}&:\,\,E_0=\frac{1}{2}\,\left(3+ \sqrt{1+4\,m^2\,L_0^2}\right)\,,\nonumber\\
\mbox{spinors, gravitino}&:\,\,E_0=\frac{1}{2}\,\left(3+ 2\,|m\,L_0|\right)\,.\nonumber\\
\end{align}

\section{$\mathcal{N}=3$ Vacua: Computational Details}
\label{app:details}
\subsection{Case ${\bf 5}\rightarrow {\bf 2}+{\bf 2}+{\bf 1}$}

The ${\rm SO}(3)_{\rm d}$-invariant  tensors $A_{ij},\,A^i{}_{jkl}$ have, in general,  the following non-vanishing components:
\begin{align}
A_{\alpha\beta}&=\delta_{\alpha\beta}\,\,;\,\,\,A_{\Lambda\Sigma}=A_{77}\,\delta_{\Lambda\Sigma}\,\,;\,\,\,A_{88}\,,\nonumber\\
A^\Lambda{}_{\Sigma\alpha\beta}&=A^{(0)}\,\epsilon_{\alpha\beta\gamma}\,(t^{(-)}_{\gamma})^\Lambda{}_{\Sigma}+
A^{(\hat{\alpha})}\,\epsilon_{\alpha\beta\gamma}\,(t^{(-)}_{\gamma}t^{(+)}_{\hat{\alpha}})^\Lambda{}_{\Sigma}\,,\nonumber\\
A^\Lambda{}_{\Sigma\alpha 8}&=C^{(0)}\,(t^{(-)}_{\alpha})^\Lambda{}_{\Sigma}+
C^{(\hat{\alpha})}\,(t^{(-)}_{\alpha}t^{(+)}_{\hat{\alpha}})^\Lambda{}_{\Sigma}\,,\nonumber\\
A^\Lambda{}_{\Sigma\Gamma\Delta}&=D^{(0)}\,\epsilon_{\Lambda\Sigma\Gamma\Delta}\,+
D^{(\hat{\alpha})}\,\delta^\Lambda_{[\Sigma}(t^{(+)}_{\hat{\alpha}})_{\Gamma\Delta]}\,,\nonumber\\
A^8{}_{\alpha\beta\gamma}&=B^{(0)}\,\epsilon_{\alpha\beta\gamma}\,\,;\,\,\,A^8{}_{8\Lambda\Sigma}=B^{(\hat{\alpha})}\,
(t^{(+)}_{\hat{\alpha}})_{\Lambda\Sigma}\,\,;\,\,\,A^8{}_{\alpha\Lambda\Sigma}=E^{(0)}\,t^{(-)}_{\alpha \,\Lambda\Sigma}\,.
\end{align}
The traceless condition on $A^i{}_{jkl}$ sets $B^{(\hat{\alpha})}=-2 \, D^{(\hat{\alpha})}/3$, so that we end up with the 16 independent parameters:
\begin{equation}
A_{77},\,A_{88},\,A^{(0)},\,A^{(\hat{\alpha})},\,B^{(0)},\,C^{(0)},\,C^{(\hat{\alpha})},\,D^{(0)},\,D^{(\hat{\alpha})},\,E^{(0)}\,.
\end{equation}
Using the residual ${\rm SU}(8)$ symmetry we can set $A_{77}$ to be real. It is useful to identify the last 14 parameters with entries of $A_i{}^{jkl}$:
\begin{align}
A^{(0)}&=2\,A^7{}_{236}\,\,;\,\,\,A^{(1)}=4\,A^7{}_{237}\,\,;\,\,\,A^{(2)}=4\,A^7{}_{234}\,\,;\,\,\,
A^{(3)}=-4\,A^7{}_{235}\,,\nonumber\\
C^{(0)}&=-2\,A^7{}_{348}\,\,;\,\,\,C^{(1)}=-4\,A^7{}_{358}\,\,;\,\,\,C^{(2)}=4\,A^7{}_{368}\,\,;\,\,\,
C^{(3)}=-4\,A^7{}_{378}\,,\nonumber\\
D^{(0)}&=-A^7{}_{456}\,\,;\,\,\,D^{(1)}=-3\,A^8{}_{678}\,\,;\,\,\,D^{(2)}=-3\,A^8{}_{578}\,\,;\,\,\,
D^{(3)}=-3\,A^8{}_{568}\,,\nonumber\\
B^{(0)}&=A^8{}_{123}\,,\,\;\,\,\,E^{(0)}=2\,A^8{}_{356}\,.
\end{align}
Aside from the known ${\cal N}=8$ solution $A_i{}^{jkl}=0,\,A_{77}=\pm 1,\,\,A_{88}=e^{i\varphi}$, we only find the  the following ${\cal N}=3$ solution (\ref{s5221}),
corresponding to $A^{(\hat{\alpha})}=D^{(\hat{\alpha})}=C^{(\hat{\alpha})}=0$, $B^{(0)}=0$ and
\bea
&&{} A_{77}=\frac{3}{2}\,\epsilon\,\,;\,\,\,A_{88}=-\sqrt{3}\,e^{2i\varphi}\,\,;\,\,\,A^{(0)}=1\,\,;\,\,\,
C^{(0)}=-\sqrt{3}\,e^{i\varphi}\,,\nonumber\\
&&{}
D^{(0)}=\epsilon \frac{\sqrt{3}}{2}\,e^{-i\varphi}\,\,;\,\,\,E^{(0)}=-2\,
\epsilon\,e^{-2i\varphi}\,.
\eea

\subsection{Case ${\bf 5}\rightarrow {\bf 3}+{\bf 1}+{\bf 1}$}
Let us split the index $i$ into $i=\alpha,\alpha',a$, where $\alpha=1,2,3$, $\alpha'=4,5,6$ and $a=7,8$ is the index labeling the singlets.
The ${\rm SO}(3)_{\rm d}$ generators in the ${\bf 8}$ of ${\rm SO}(8)$ read:
\begin{equation}
t_\alpha=\left(\begin{matrix}\epsilon_{\beta\alpha\gamma} & {\bf 0} & {\bf 0}\cr {\bf 0} & \epsilon_{\beta'\alpha\gamma'} & {\bf 0}\cr {\bf 0}& {\bf 0}& {\bf 0}_2\end{matrix}\right)\,,
\end{equation}
and satisfy the relations (\ref{so3alg}).\par
The ${\rm SO}(3)_{\rm d}$-invariant tensors $A_{ij},\,A^i{}_{jkl}$ have the following non-vanishing components:
\begin{align}
A_{\alpha\beta}&=\delta_{\alpha\beta}\,\,;\,\,\,A_{\alpha'\beta'}=A_{66}\,\delta_{\alpha'\beta'}\,\,;\,\,\,A_{ab}\,,\nonumber\\
A^a{}_{\alpha'\beta'\gamma'}&=A^a\,\epsilon_{\alpha'-3\,\beta'-3\,\gamma'-3}\,\,;\,\,\,A^a{}_{\alpha'\beta'\gamma}=B^a\,
\epsilon_{\alpha'-3\,\beta'-3\,\gamma}\,,\nonumber\\
A^a{}_{\alpha'\beta\gamma}&=D^a\,\epsilon_{\alpha'-3\,\beta\gamma}\,\,;\,\,\,A^a{}_{\alpha\beta\gamma}=C^a\,
\epsilon_{\alpha\beta\gamma}\,,\nonumber\\
A^a{}_{b'\beta'\gamma}&= A^a{}_b\,\delta_{\beta'-3\,\gamma}\,\nonumber\\
A^{\alpha'}{}_{\beta'\gamma' a}&=\tilde{A}_a\,\epsilon_{\alpha'-3\,\beta'-3\,\gamma'-3}\,\,;\,\,\,A^{\alpha'}{}_{\beta'\gamma a}=\tilde{B}_a\,\epsilon_{\alpha'-3\,\beta'-3\,\gamma}\,\,;\,\,\,A^{\alpha'}{}_{\beta\gamma a}=\tilde{D}_a\,\epsilon_{\alpha'-3\,\beta\,\gamma}\,,\nonumber\\
A^{\alpha'}{}_{\alpha ab}&=B\,\delta^{\alpha'-3}_\alpha\,\epsilon_{ab}\,\,\,;\,\,\,\,A^{\alpha'}{}_{\beta' \alpha\beta}=E\,\delta^{\alpha'-3\,\beta'-3 }_{\alpha\beta}\,\,\,;\,\,\,\,A^{\alpha'}{}_{\beta \alpha\gamma}=C\,\delta^{\alpha'-3\,\beta }_{\alpha\gamma}\,.
\end{align}
We can always set $A_{78}=0$ and $A_{66},\,A_{77}$ to be real and the 21 complex parameters entering $A^i{}_{ijk}$:
\begin{equation}
A^a{}_b,\,A^a,\,B^a,\,C^a,\,D^a,\,\tilde A^a,\,\tilde B^a\,\tilde D^a,\, B,\, C,\,E\,,\label{parA2}
\end{equation}
are subject to the tracelessness condition:
\begin{equation}
C=A^a{}_a\,,
\end{equation}
which leaves us with a total of 23 complex parameters $$A_{66},\,A_{aa},\,A^a{}_b,\,A^a,\,B^a,\,C^a,\,D^a,\,\tilde A^a,\,\tilde B^a\,\tilde D^a,\, B,\, C,\,E\,,$$
two of which ($A_{66},\,A_{77}$), as previously mentioned, can be made real. The relation of the parameters (\ref{parA2}) to the entries of $A^i{}_{ijk}$ is:
\begin{align}
A^a{}_b&=-A^a{}_{b 36}\,\,;\,\,\,A^a=A^a{}_{456}\,\,;\,\,\,B^a=A^a{}_{345}\,\,;\,\,\,C^a=A^a{}_{123}\,,\nonumber\\
D^a&=A^a{}_{234}\,\,;\,\,\,\tilde A_a=A^6{}_{45a}\,\,;\,\,\,\tilde B_a=-A^6{}_{24a}\,\,;\,\,\,\tilde D_a=A^6{}_{12a}\,,\nonumber\\
B&=A^6{}_{378}\,\,;\,\,\,E=-2\,A^6{}_{235}\,.
\end{align}
Aside from the known ${\cal N}=8$ solution , we find the following solution
\begin{align}
A_{66}&=2\,\xi\,\,\,;\,\,\,A_{77}=2\,\eta\,\,;\,\,\,A_{88}=\xi\,\eta\,e^{i\varphi}\,,\nonumber\\
A^a&=C^a=D^a=0\,\,\,;\,\,\,B^7=\sqrt{2}\,\xi\,\eta\,e^{-i\frac{\varphi}{4}}\,\,\,;\,\,\,B^8=0\,,\nonumber\\
\tilde A_7&=0\,\,\,;\,\,\,\tilde A_8=\sqrt{2}\,\xi\,e^{i\frac{\varphi}{4}}\,\,\,;\,\,\,\tilde B_7=-\sqrt{2}\,e^{-i\frac{\varphi}{4}}\,\,\,;\,\,\,\tilde B_8=0\,\,\,;\,\,\,\tilde D_a=0\,,\nonumber\\
C&=0\,\,;\,\,\,B=-\eta\,e^{i\frac{\varphi}{2}}\,\,;\,\,\,E=-2\,,\nonumber\\
A^7{}_7&=A^8{}_8=A^8{}_7=0\,\,;\,\,\,A^7{}_8=-\xi\,e^{i\frac{\varphi}{2}}\,,
\end{align}
where $\eta,\,\xi=\pm 1$. Using the above identifications we can write the non-vanishing components of  the tensors as in (\ref{s53111}).
%%%%%%%%%%%%%%%%%%%%%%%%%%%%%%%%
%%%%%%%%%%%%%%%%%%%%%%%%%%%%%%%%

%\bibliographystyle{JHEP2}
%\bibliography{refs}

\begin{thebibliography}{10}

\bibitem{deWit:1982ig}
B.~de~Wit and H.~Nicolai, { ${N}=8$ supergravity},  { Nucl. Phys.} { B208}
  (1982)
323.
%%CITATION = NUPHA,B208,323;%%.

\bibitem{Warner:1983vz}
N.~P. Warner, { Some new extrema of the scalar potential of gauged ${N=8}$
  supergravity},  { Phys. Lett.} { B128} (1983)
169.
%%CITATION = PHLTA,B128,169;%%.

\bibitem{Fischbacher:2008zu}
T.~Fischbacher, { The many vacua of gauged extended supergravities},  {
  Gen.Rel.Grav.} { 41} (2009) 315--411,
[\href{http://xxx.lanl.gov/abs/0811.1915}{{\tt 0811.1915}}].
%%CITATION = 0811.1915;%%.

\bibitem{Hull:1984qz}
C.~M. Hull, { More gaugings of ${N}=8$ supergravity},  { Phys. Lett.} { B148}
  (1984)
297--300.
%%CITATION = PHLTA,B148,297;%%.

\bibitem{Hull:1988jw}
C.~M. Hull and N.~P. Warner, { Noncompact gaugings from higher dimensions},  {
  Class. Quant. Grav.} { 5} (1988)
1517.
%%CITATION = CQGRD,5,1517;%%.

\bibitem{Andrianopoli:2002mf}
L.~Andrianopoli, R.~D'Auria, S.~Ferrara, and M.~A. Lledo, { Gauging of flat
  groups in four dimensional supergravity},  { JHEP} { 07} (2002) 010,
[\href{http://xxx.lanl.gov/abs/hep-th/0203206}{{\tt hep-th/0203206}}].
%%CITATION = HEP-TH/0203206;%%.

\bibitem{deWit:2002vt}
B.~de~Wit, H.~Samtleben, and M.~Trigiante, { On {L}agrangians and gaugings of
  maximal supergravities},  { Nucl. Phys.} { B655} (2003) 93--126,
[\href{http://xxx.lanl.gov/abs/hep-th/0212239}{{\tt hep-th/0212239}}].
%%CITATION = HEP-TH 0212239;%%.

\bibitem{deWit:2005ub}
B.~de~Wit, H.~Samtleben, and M.~Trigiante, { {Magnetic charges in local field
  theory}},  { JHEP} { 09} (2005) 016,
[\href{http://xxx.lanl.gov/abs/hep-th/0507289}{{\tt hep-th/0507289}}].
%%CITATION = HEP-TH/0507289;%%.

\bibitem{deWit:2007mt}
B.~de~Wit, H.~Samtleben, and M.~Trigiante, { {The maximal ${D} = 4$
  supergravities}},  { JHEP} { 06} (2007) 049,
[\href{http://xxx.lanl.gov/abs/arXiv:0705.2101}{{\tt arXiv:0705.2101}}].
%%CITATION = ARXIV:0705.2101;%%.

\bibitem{Dibitetto:2011gm}
G.~Dibitetto, A.~Guarino, and D.~Roest, { Charting the landscape of {${\cal
  N}=4$} flux compactifications},  { JHEP} { 1103} (2011) 137,
[\href{http://xxx.lanl.gov/abs/1102.0239}{{\tt 1102.0239}}].
%%CITATION = ARXIV:1102.0239;%%.

\bibitem{DallAgata:2011aa}
G.~Dall'Agata and G.~Inverso, { On the vacua of {$N = 8$} gauged supergravity
  in 4 dimensions},  { Nucl.Phys.} { B859} (2012) 70--95,
[\href{http://xxx.lanl.gov/abs/1112.3345}{{\tt 1112.3345}}].
%%CITATION = ARXIV:1112.3345;%%.

\bibitem{DallAgata:2012bb}
G.~Dall'Agata, G.~Inverso, and M.~Trigiante, { Evidence for a family of
  ${SO}(8)$ gauged supergravity theories},  { Phys.Rev.Lett.} { 109} (2012)
  201301,
[\href{http://xxx.lanl.gov/abs/1209.0760}{{\tt 1209.0760}}].
%%CITATION = ARXIV:1209.0760;%%.

\bibitem{DallAgata:2012sx}
G.~Dall'Agata and G.~Inverso, { de {S}itter vacua in {$N = 8$} supergravity and
  slow-roll conditions},  { Phys.Lett.} { B718} (2013) 1132--1136,
[\href{http://xxx.lanl.gov/abs/1211.3414}{{\tt 1211.3414}}].
%%CITATION = ARXIV:1211.3414;%%.

\bibitem{DallAgata:2014ita}
G.~Dall'Agata, G.~Inverso, and A.~Marrani, { Symplectic deformations of gauged
  maximal supergravity},  { JHEP} { 1407} (2014) 133,
[\href{http://xxx.lanl.gov/abs/1405.2437}{{\tt 1405.2437}}].
%%CITATION = ARXIV:1405.2437;%%.

\bibitem{Borghese:2012qm}
A.~Borghese, A.~Guarino, and D.~Roest, { All ${G}_2$ invariant critical points
  of maximal supergravity},  { JHEP} { 1212} (2012) 108,
[\href{http://xxx.lanl.gov/abs/1209.3003}{{\tt 1209.3003}}].
%%CITATION = ARXIV:1209.3003;%%.

\bibitem{Borghese:2012zs}
A.~Borghese, G.~Dibitetto, A.~Guarino, D.~Roest, and O.~Varela, { The
  {$SU(3)$}-invariant sector of new maximal supergravity},  { JHEP} { 1303}
  (2013) 082,
[\href{http://xxx.lanl.gov/abs/1211.5335}{{\tt 1211.5335}}].
%%CITATION = ARXIV:1211.5335;%%.

\bibitem{Borghese:2013dja}
A.~Borghese, A.~Guarino, and D.~Roest, { Triality, periodicity and stability of
  {$SO(8)$} gauged supergravity},  { JHEP} { 1305} (2013) 107,
[\href{http://xxx.lanl.gov/abs/1302.6057}{{\tt 1302.6057}}].
%%CITATION = ARXIV:1302.6057;%%.

\bibitem{LeDiffon:2011wt}
A.~Le~Diffon, H.~Samtleben, and M.~Trigiante, { {$N=8$} supergravity with local
  scaling symmetry},  { JHEP} { 1104} (2011) 079,
[\href{http://xxx.lanl.gov/abs/1103.2785}{{\tt 1103.2785}}].
%%CITATION = ARXIV:1103.2785;%%.

\bibitem{Andrianopoli:2008ea}
L.~Andrianopoli, R.~D'Auria, S.~Ferrara, P.~Grassi, and M.~Trigiante, {
  Exceptional ${N}=6$ and ${N}=2$ {AdS}$_4$ supergravity, and zero-center
  modules},  { JHEP} { 0904} (2009) 074,
[\href{http://xxx.lanl.gov/abs/0810.1214}{{\tt 0810.1214}}].
%%CITATION = ARXIV:0810.1214;%%.

\bibitem{Freedman:1983na}
D.~Z. Freedman and H.~Nicolai, { Multiplet shortening in {Osp($N$,4)}},  {
  Nucl.Phys.} { B237} (1984)
342.
%%CITATION = NUPHA,B237,342;%%.

\bibitem{Termonia:1999cs}
P.~Termonia, { The complete ${{\cal N}}=3$ {K}aluza-{K}lein spectrum of $11{D}$
  supergravity on {$AdS_4 \times N^{0,1,0}$}},  { Nucl.Phys.} { B577} (2000)
  341--389,
[\href{http://xxx.lanl.gov/abs/hep-th/9909137}{{\tt hep-th/9909137}}].
%%CITATION = HEP-TH/9909137;%%.

\bibitem{Fre:1999xp}
P.~Fr\'e, L.~Gualtieri, and P.~Termonia, { The structure of {${\cal N}=3$}
  multiplets in {$AdS_4$} and the complete {$Osp(3|4) \times SU(3)$} spectrum
  of {M} theory on {$AdS_4 \times N^{0,1,0}$}},  { Phys.Lett.} { B471} (1999)
  27--38,
[\href{http://xxx.lanl.gov/abs/hep-th/9909188}{{\tt hep-th/9909188}}].
%%CITATION = HEP-TH/9909188;%%.

\bibitem{Breitenlohner:1982jf}
P.~Breitenlohner and D.~Z. Freedman, { Stability in gauged extended
  supergravity},  { Annals Phys.} { 144} (1982)
249.
%%CITATION = APNYA,144,249;%%.

\bibitem{Aharony:2008gk}
O.~Aharony, O.~Bergman, and D.~L. Jafferis, { Fractional {M2}-branes},  { JHEP}
  { 11} (2008) 043,
[\href{http://xxx.lanl.gov/abs/0807.4924}{{\tt 0807.4924}}].
%%CITATION = 0807.4924;%%.

\bibitem{Maldacena:1997re}
J.~M. Maldacena, { The large ${N}$ limit of superconformal field theories and
  supergravity},  { Adv. Theor. Math. Phys.} { 2} (1998) 231--252,
[\href{http://xxx.lanl.gov/abs/hep-th/9711200}{{\tt hep-th/9711200}}].
%%CITATION = HEP-TH 9711200;%%.
\bibitem{Guarino:2013gsa}
  A.~Guarino,
  ``On new maximal supergravity and its BPS domain-walls,''
  JHEP {\bf 1402} (2014) 026
  [arXiv:1311.0785 [hep-th]];
  %%CITATION = ARXIV:1311.0785;%%
\bibitem{Tarrio:2013qga}
J.~Tarr{\'\i}o and O.~Varela, { Electric/magnetic duality and {RG} flows in
  {AdS$_4$/CFT$_3$}},  { JHEP} { 1401} (2014) 071,
[\href{http://xxx.lanl.gov/abs/1311.2933}{{\tt 1311.2933}}].
%%CITATION = ARXIV:1311.2933;%%.

\bibitem{Anabalon:2013eaa}
A.~Anabalon and D.~Astefanesei, { {Black holes in $\omega$-defomed gauged $N=8$
  supergravity}},  { Phys.Lett.} { B732} (2014) 137--141,
[\href{http://xxx.lanl.gov/abs/1311.7459}{{\tt 1311.7459}}].
%%CITATION = ARXIV:1311.7459;%%.

\bibitem{Freedman:1999gp}
D.~Z. Freedman, S.~S. Gubser, K.~Pilch, and N.~P. Warner, { {Renormalization
  group flows from holography supersymmetry and a $c$-theorem}},  { Adv. Theor.
  Math. Phys.} { 3} (1999) 363--417,
[\href{http://xxx.lanl.gov/abs/hep-th/9904017}{{\tt hep-th/9904017}}].
%%CITATION = HEP-TH/9904017;%%.

\end{thebibliography}

\providecommand{\href}[2]{#2}\begingroup\raggedright\endgroup

\end{document}